\documentclass[a4paper,12pt]{article}
\usepackage[margin=1.2in,bottom=1in,top=1in]{geometry}

\usepackage{graphicx} 
\usepackage{ragged2e} 
\usepackage{comment} 
\usepackage{enumitem}
\usepackage{caption}
\usepackage[table,xcdraw]{xcolor}
\usepackage{hyperref}
\hypersetup{
colorlinks=true,
urlcolor= blue,
citecolor=blue,
linkcolor= black
}
\usepackage{braket}
\usepackage{framed}
\usepackage{soul}
\usepackage{xcolor}
\definecolor{blue}{RGB}{33, 118, 199}

\definecolor{green}{RGB}{0, 128, 0}

\definecolor{red}{RGB}{230, 0, 20}

\usepackage{tikz}
\usepackage{amsmath}
\usepackage{amssymb}
\usepackage{comment}
\usepackage{dsfont}
\usepackage{xcolor}
\usepackage{cancel}
\usepackage{longtable}
\usepackage{empheq}
\usepackage{multirow}

\newcommand{\R}{\mathbb{R}}
\newcommand{\Z}{\mathbb{Z}}

\newcommand{\normord}[1]{:\mathrel{#1}:}

\usepackage[T1]{fontenc}
\usepackage{fancyhdr}

\begin{document}

\begin{titlepage}
\hfill \hbox{\normalsize KEK-TH-2552
}\\
\vspace*{15mm}
\begin{center}
\begin{FlushLeft}
{ \Large \sffamily  \textbf{On the String Landscape Without Hypermultiplets} }

\rule{\linewidth}{1pt}
\vspace*{5mm}
\vspace{1cm}
{
Zihni Kaan Baykara$^1$, Yuta Hamada$^{2,3}$, Houri-Christina Tarazi{$^4$}, Cumrun Vafa$^1$\\}
\vspace{.6cm}
{ \itshape $^1$ Department of Physics, Harvard University, Cambridge, Massachusetts, USA}\par
\vspace{.2cm}
{\itshape $^2$ Theory Center, IPNS, High Energy Accelerator Research Organization (KEK), 1-1 Oho, Tsukuba, Ibaraki 305-0801, Japan}\par
\vspace{.2cm}
{\itshape $^3$ Graduate Institute for Advanced Studies, SOKENDAI, 1-1 Oho, Tsukuba, \\Ibaraki 305-0801, Japan}\par\vspace{.2cm}
{\itshape $^4$ Enrico Fermi Institute \& Kadanoff Center for Theoretical Physics, University of Chicago, Chicago, IL 60637, USA}\par
\vspace{.2cm}\par
\vspace{-.3cm}

\vspace{1cm}
\begin{abstract}
In this work we study interesting corners of  the quantum gravity  landscape with 8 supercharges pushing the boundaries of our current understanding.  Calabi-Yau threefolds compactifications of F/M/type II theories to 6, 5 and 4 dimensions are the most prominent examples of this class, and these always lead to a universal hypermultiplet coming from the volume/string coupling constant.
We find that there are asymmetric orbifold constructions which have no hypermultiplets in 4 or 5 dimensions and no neutral hypers in 6d.
We argue that these theories can also be obtained by going to strong coupling/small volume regions of geometric constructions where a new Coulomb branch opens up and moving in this direction freezes the volume/string coupling constant. Interestingly we find that the Kodaira condition encountered in geometric limits of F-theory compactifications to 6 dimensions is violated in these corners of the landscape due to strong quantum corrections. We also construct a theory in 3 dimensions which if it were to arise by toroidal compactifications from 5d, it would have to come from pure ${\cal N}=1$ supergravity with no massless scalar fields.

\end{abstract}
\end{FlushLeft}
\end{center}
\end{titlepage}

\tableofcontents

\newpage

\section{Introduction}
The Swampland program~\cite{Vafa:2005ui} has used the string landscape as the guiding principle to determine what are some universality features of quantum gravitational theories (see \cite{Palti:2019pca,vanBeest:2021lhn,Grana:2021zvf,Agmon:2022thq} for a review).  As such it is crucial to try to characterize exotic corners of quantum gravity(QG) landscape, so we can uncover sharper Swampland criteria.

The study of supersymmetric theories with 32 or more non-trivially 16 supercharges in Minkowski space has advanced quite a bit and by now we can have a bottom up derivation~\cite{Kim:2019ths,Montero:2020icj,Cvetic:2020kuw,Hamada:2021bbz,Bedroya:2021fbu} of allowed possibilities for  low energy EFT's based on Swampland principles, which matches what we know from the string landscape for $d\geq 7$~\cite{deBoer:2001wca,Aharony:2007du,Font:2020rsk,Font:2021uyw,Fraiman:2021soq,Cvetic:2022uuu,ParraDeFreitas:2022wnz,Montero:2022vva}.  Beginning in $d=6$ we also have the possibility of having 8 supercharges:
${\cal N}=(1,0)$ in $d=6$, ${\cal N}=1$ in $d=5$, ${\cal N}=2$ in $d=4$ and ${\cal N}=4$ in $d=3$.  It is important to comprehensively study the known string landscape for this class as well.

The most straight-forward way of constructing this class of theories involves compactification of F-theory, M-theory and type II theories on CY 3-folds.  There is one commonality in all these geometric constructions of the string landscape:  We always have a universal hypermultiplet, which corresponds to volume of the space/string coupling.  It would be important to understand whether the universal hypermultiplet is an unavoidable feature of consistent QG or an artifact of geometric compactification.  In fact, it is already known that there are non-geometric string compactifications which lead to no hypermultiplets~\cite{Kachru:1995wm}, more recently \cite{Israel:2013wwa,Hull:2017llx,Gkountoumis:2023fym}.  We perform a more exhaustive search of such compactifications by studying asymmetric orbifolds~\cite{Narain:1986qm,Narain:1990mw} of heterotic and type II theories.  Moreover, we extend this class to 6d which leads to the construction of models without any neutral hypermultiplets.  In doing so we discover a number of interesting features:  We construct a number of theories in 6d which have no neutral hypers, but when we higgs the charged ones we land on a geometric compactification of F-theory  on elliptic 3-fold~\cite{Vafa:1996xn,Morrison:1996na,Morrison:1996pp}. Since the geometric limit of F-theory, with large volume for the base of the elliptic 3-fold, always has a neutral multiplet,  this implies that when the volume of the base is Planckian the volume modulus itself becomes part of the charged multiplet (similar features were noted in an example in \cite{Kachru:1995wm} in 4d).  Upon going down to 5 and 4 dimensions this leads to theories which have no hypermultiplets at all, by going to the Coulomb branch which masses up the charged hypers.  In doing so, we find an interesting structure in the rank of the 4d and 5d gauge groups in 8 supercharge theories with no hypers:
Firstly, we find that the rank is  always less than or equal to $r_G\leq (26-d)+1$.  Secondly, we only find even ranks in 5d and odd ones in 4d.  We do not have a bottom up explanation of these features. An additional interesting result is about the Kodaira condition \cite{Kumar_2010} which holds for all F-theory constructions. In particular, we show that this condition can be violated for non-geometric models and therefore it is not a feature of 6d supersymmetric quantum gravity.

A more non-trivial question is whether one can have a pure ${\cal N}=1$ supergravity in 5d with no additional fields.   This would not be easy to construct from string theory directly as it would have no weak coupling limits.  However, it is conceivable that it would have been discovered by its compactifications to lower dimensions, as in the connection between M-theory in 11 dimensions and type IIA in 10d \cite{Mizoguchi:2001cp}.  We have not been able to construct such a theory in 4d (which would have required a single vector multiplet).  However, we have been able to construct a theory in 3d which has the same multiplet as that of the 5d pure supergravity theory compactified on $T^2$.  It remains to show whether this has a 5d uplift.

The organization of this paper is as follows: In Section~\ref{sec:6d}, we provide 6d models without neutral hypermultiplets based on the asymmetric orbifolds. We find that the Kodaira condition is not satisfied in this class of models. Many 5d and 4d models without hypermultiplets are constructed in Section~\ref{sec:5d} and Section~\ref{sec:4d}, respectively. We conclude in Section~\ref{sec:conlusion}.  Technical details are discussed in the appendices.

\section{6d models without neutral hypermultiplets}\label{sec:6d}

The highest dimension where we can have a theory with 8 supercharges is in $d=6$ where  the minimal ${\cal N}=(1,0)$ supersymmetry arises.
The landscape of 6d minimally supersymmetric supergravities has been of particular interest due to the very constraining chiral anomalies which significantly reduce the possible theories.

A large class of models can be constructed using F-theory~\cite{Vafa:1996xn,Morrison:1996na,Morrison:1996pp} compactified on elliptic Calabi-Yau threefolds. The web of string dualities provides also descriptions in terms of the heterotic string on K3, M-theory on $K3\times S^1/\mathbb{Z}_n$ or type II orientifold/ brane constructions. However, these constructions correspond to a subset of the string landscape and anomalies seem to indicate that more possibilities can be allowed.

In particular, the gravitational anomaly for the 6d $\mathcal{N}=1$ theory is given by 
\begin{eqnarray}
    H-V=273-29T
\end{eqnarray}
where $H,T,V$ are the number of hyper multiplets, tensor multiplets and vector multiplets respectively. This anomaly can be a useful indicator for the possible boundaries of the string landscape. Some of the interesting cases can be summarized as:

\begin{itemize}
    \item $T=0$: Theories without any tensor multiplets have been studied in \cite{Kumar:2010am} and have  constructions given by F-theory on an elliptic threefold with base $\mathbb{P}_2$ \cite{Morrison:1996pp} or through type I or Gepner models in \cite{Angelantonj:1996mw}. One can also arrive to these models through extremal transitions from $T>0$ as described in \cite{Morrison:1996pp}.
    \item $H_{charged}=0, \ H_{neutral}=0,\ H_{charged}+H_{neutral}=0$: Here we need to distinguish between the following  cases:
    \begin{itemize}
        \item 
        $H_{charged}=0$: These are theories with no charged hypermultiplets, which are the typical F-theory examples in the maximally Higgsed phases. For example the bases $\mathbb{P}^2,\mathbb{F}_{n=0,2,3,4,5,6,8,12}$ all have no charged matter.\cite{Morrison:2012np, Morrison:1996pp,Seiberg:1996vs}.

        \item 
        $H_{neutral}=0$: This is a very interesting case as described in the beginning of the section because no geometric construction can give rise to such a theory as the volume element always provides a free neutral hyper in the geometric limits of Type II or F/M-theory pictures. Such constructions (and the fate of the universal hypermultiplet in some strong coupling limits) are the main goal of this section and a number of examples will be developed together with their relations to geometric models. Examples of these models have appeared previous in the literature including orbifold constructions \cite{Kachru:1995wm, Gkountoumis:2023fym}, Gepner models \cite{Israel:2013wwa} and free fermion constructions \cite{Dolivet:2007sz}.       
        \item $H=0$: In this case the gravitational anomaly implies that $V=29 T-273$ and hence $T\geq 10$. 
In terms of orbifold constructions such a theory cannot come from a perturbative heterotic string as those cases have only $T=1$. In the type II picture one could potentially realize such a theory but no construction is known.\footnote{It could have only abelian gauge factors or some combination with non-abelian enhancements. Any non-abelian gauge symmetry for the type II string comes from the NS-NS sector and contributes to $c_L=4$. If one is looking for non-abelian gauge symmetries with no hypermultiplets then the only possibilities are the Non-Higgsable clusters and the ones with $\sum_i c_{G_i}\leq 4$ are $SU(3)_1\times SU(3)_1,SU(3)_{k\leq 3},SO(8)_1$ by unitarity. }.
        
    \end{itemize}
    \item $V=0$: Such models typically occur when no Non-Higgsable clusters are present in  F-theory constructions as for example those  with base $\mathbb{F}_n$ where $n=0,1,2$ or $\mathbb{P}^2$.
\end{itemize}

We can also consider pairs of these conditions: $V=T=0$ can be realized as the generic model for F-theory with bases $\mathbb{P}^2$. The $H=T=0$ model would require $V=-273$ which is thus ruled out. 

If however we considered the case $H_{neutral}=T=0$ then we would require $H_{charged}-V=273$ which can have non-abelian groups that satisfy anomalies and unitarity as shown in \cite{Kim:2019vuc}. Any such theory does not have neutral scalar fields so it does not have a perturbative string construction but it could potentially appear as the strong coupling limit of a model where some neutral fields become charged. For example, if the model is Higgsable and it is possible to tune all neutral hypers so that they become charged similar to other examples we study here then this could be realized. An example, could be an F-theory model on $\mathbb{P}^2$ where the moduli controlling the base volume also gets charged. It is hard to argue for the existence of such special points but they are not ruled out.  Finally, note that the pure $\mathcal{N}=(1,0)$ supergravity with $H=V=0$  is not allowed as it violates anomalies.

The most interesting of the above cases is $H=0$ because these cannot be constructed using geometric constructions but they could potentially be related to them through some transitions.
A large class of non-geometric examples  can be constructed using  asymmetric orbifolds \cite{Narain:1986qm}  which can provide new classes of theories not seen in normal geometric constructions. Such constructions as summarized in \autoref{sec:review} and will be the model building tool of this paper.

As mentioned above usual geometric models always have a free \textit{universal hypermultiplet } (in the type II picture) associated to the volume element of the manifold, but one could consider a process where this scalar can become charged at some special points where the volume freezes. Indeed models without hypermultiplets can be realized in asymmetric orbifolds \cite{Kachru:1995wm} in 4d.  Interestingly, in the same work it was conjectured that in fact such models are related to geometric models through strong coupling analogs which involve going to the small volume limit of the Calabi-Yau manifold and hence the geometry semi-classical description is no longer valid.  These are quantum versions of conifold transitions involving higgsing/unhiggsing the vectors. Such transitions are particularly interesting because they provide a fresh look at the connectedness of string vacua and  provide examples of strong coupling dualities. The loss of a geometric picture on the type II side makes it harder to understand these transitions precisely but there are good indications, including duality with heterotic strings where these enhanced gauge symmetries become pertubatively manifest \cite{Kachru:1995wm} that they should in fact be possible. 

Here we study asymmetric orbifolds of both Type II and heterotic string theories that have no neutral hypers and propose dual geometric string/M-theory vacua involving compactifications on geometric Calabi-Yau manifolds that they could be related to through Higgsing/unHiggsing similar to the ones described above.

Another interesting observation is that certain geometric consistency condition is violated in some non-geometric models. 
In particular, as summarized in \autoref{app:6dreview} an ${\cal N}=(1,0)$ 6d theory is specified by vectors $-a, b_i\in \mathbb{R}^{1,T}$corresponding to the  gravitational and gauge instantonic string charges. In the F-theory picture these charges correspond to the canonical class and effective curves wrapped by D3 in the base of the elliptic threefold.
A consistent elliptic fibration in F-theory needs to satisfy the\textit{ Kodaira condition} \cite{Kumar_2010} given by 
\begin{equation}\label{eq:kodaira}
    -12a=\sum_i \nu_i b_i+Y
\end{equation}
where $\nu_i$ are the singularity multiplicities and $Y$ the residual divisors which is the sum of effective divisors associated to non-contractible curves or  abelian gauge factors. From the bottom up supergravity perspective the divisors $Y$ correspond to a collection of supergravity strings \footnote{A supergravity string~\cite{Katz:2020ewz, Tarazi:2021duw} of charge $Q$ is a BPS string that only exists in a gravity theory and satisfies $Q^2\geq 0 $ and $Q\cdot D\geq 0$ for any BPS string of charge $D$. }. This means that  tensor branch scalars $j$ satisfy $j\cdot Y \geq 0 $ and hence 
\begin{equation}\label{eq:kodairaJ}
  - j\cdot( 12a+\sum_i \nu_i b_i) \geq 0 
\end{equation}
 
The quantities $-j\cdot  a$ and $j\cdot b_i$ correspond to the central charge in the supersymmetry algebra for tension of the strings with charges $a, \ b_i$ respectively in the EFT description. In particular, in the case that these strings correspond to gauge instantons of some gauge symmetry of the theory the tensions control the corresponding gauge coupling $g^2=T^{-1}$. Therefore, the  Kodaira constraint \autoref{eq:kodairaJ} can be phrased in purely EFT terms with no reference to F-theory.

This constraint was suspected to hold for any 6d quantum gravity with 8 supercharges.  In fact in \cite{Hamada:2021bbz}, where theories with 16 supercharges where studied a similar condition was shown to be true, corresponding to the 8d analogue of the Kodaira condition. In that work it was argued that small instantons can be exchanged with 3-brane probes which preserve 8 supercharges. According to the cobordism conjecture \cite{McNamara:2019rup} different configurations are expected to be connected and due to the amount of supersymmetry they are expected to be connected through supersymmetric deformations.  However, in this case they would preserve only 4 supercharges and even though the cobordism conjecture still implies they are connected in field space, the path could go through non-supersymmetric deformations as a superpotential is now allowed. This ended up being an obstacle in proving the Kodaira condition in lower supersymmetric cases.
 Thus the proof never materialized, and raised the question whether the Kodaira condition is generally true or not?  Interestingly here we find that this condition is not universal for 6d theories by making explicit constructions of string theories that \textit{do not} preserve \autoref{eq:kodairaJ}. To our knowledge this is the first example that demonstrates this! This also serves as a good lesson that the inability to support such constraints from bottom up was a valid motivation to doubt its validity in the context of the Swampland program and leads to searching for them in the string landscape.  Indeed the Kodaira condition is not necessary for a consistent QG!
 
In this section we demonstrate that geometric compactifications of F-theory do not encompass the entire Landscape.\footnote{This does not necessarily imply that the theory we discover is a disconnected component of the moduli space, as we will discuss later.} We achieve this by explicitly constructing 6d models without neutral hypermultiplets using non-geometric asymmetric orbifolds which violate the Kodaira condition.

\subsection*{6d Asymmetric Orbifold Models}
The general construction of asymmetric abelian orbifolds is reviewed in \autoref{sec:review} and  involves a choice of momentum lattice which is an even self-dual lattice
\begin{eqnarray}
    \Gamma^{4,4}(\mathfrak{g})=\{(p_L,p_R)|p_L\in\Lambda_W(\mathfrak{g}),p_R\in\Lambda_W(\mathfrak{g}),p_L-p_R\in\Lambda_R(\mathfrak{g})\}.
\end{eqnarray}
for some simply-laced algebra $\mathfrak{g}$.

Additionally, one needs to specify the group action.  Each group element  is labeled by twist vectors $\phi_{L,R}$, corresponding to some automorphism of the lattice $[{\rm exp}(2\pi i \phi_L),{\rm exp}(2\pi i   \phi_R)]$  and the shift vectors $v_{L,R}$. In the case of the heterotic string one needs to specify the action on the lattice which includes in addition the associated heterotic gauge factors.

There are many such choices for the lattice and orbifold actions but our objective is to preserve 8 supercharges and the resulting theory to have no neutral hypermultiplets. For example, in the  type II picture one simple way to remove all right moving supersymmetries is by twisting by $(-1)^{F_R}$, where ${F_R}$ is the right-moving fermion number. While preserving half of the left-moving supersymmetries can be accomplished by a choice of $\phi_L$ that belongs to $SU(2)$. This leads to 8 supercharges. The objective is to choose appropriate shift vectors that can lift many of the moduli with the ultimate goal of having no free neutral hypermultiplets. In  \autoref{tab:6d_modelA}  four examples of such theories are summarized with more details left for the \autoref{app:AOdetails}. As can be seen all hypermultiplets in each of these models  are charged under some of the gauge factors of the theory. 
Such constructions cannot have geometric descriptions due to the lack of a free neutral hyper in the weakly coupled type II picture for which the dilaton (and the volume moldulus) lead to massless neutral hypers. In the weakly coupled asymmetric orbifold string constructions the string coupling is instead part of a tensor multiplet. 

\newpage
\renewcommand{\baselinestretch}{1.5}
\begin{longtable}{|c|c|c|c|c|}
\hline
\begin{tabular}{c}
     Without  \\
     neutral $H_0$ 
\end{tabular}
& \cellcolor[HTML]{FFFFC7} \textbf{Model} $1$ & \cellcolor[HTML]{FFFFC7} \textbf{Model} $2$ 
\\ \hline
Type &  II & Heterotic 
\\[+0.3mm]\hline
& & \\ 
Lattice & 
$\Gamma^{4,4}(D_4)$
&
\begin{tabular}{c}
     \\[-4mm]
     $\Gamma^{4,4}(D_4)+2\Gamma^{8,0}(E_8)$   \\[+0.9mm]
\end{tabular}
\\[+0.3mm]\hline
Twist &
\begin{tabular}{c}
     \\[-4mm]
     $\phi_L=(\frac{1}{2},\frac{1}{2}),$  \\[+0.7mm]
     $\phi_R=(1,0)$  \\[+0.9mm]
\end{tabular}
&
\begin{tabular}{c}
     \\[-4mm]
     $\phi_L=(0,0),$  \\[+0.7mm]
     $\phi_R=(\frac{1}{2},\frac{1}{2})$  \\[+0.9mm]
\end{tabular}
\\[+1.7mm]\hline
& & \\ 
Shift & No shift & $V_L=\frac{1}{2}(1^2,0^6;1^2,0^6)$
\\[+1.7mm] \hline
& & \\ 
Gauge  & $U(1)^{12}$ & {\small$E_7\times SU(2)\times E_7\times SU(2)\times SO(8)$}
\\[+1.3mm]\hline
\begin{tabular}{c}
  \\
Representation
\end{tabular}
&
\begin{tabular}{c}
    {\small$\left(\underline{\pm1,0,0,0},0^8\right)
    +\left(\pm,\pm,\pm,\pm,0^8\right)$} \\
    {\small$+\left(\underline{\pm,\mp,\mp,\mp},0^8\right)
    +\left(\underline{-,-,+,+},0^8\right)$}\\ [+0.9mm]
\end{tabular}
& 
\begin{tabular}{c}
     \\[-4mm]
     {\small $(\mathbf{56},\mathbf{2},\mathbf{1},\mathbf{1},\mathbf{1})+(\mathbf{1},\mathbf{1},\mathbf{56},\mathbf{2},\mathbf{1})$}\\
     {\small $(\mathbf{56},\mathbf{1},\mathbf{1},\mathbf{2},\mathbf{1})+(\mathbf{1},\mathbf{2},\mathbf{56},\mathbf{1},\mathbf{1})$}\\
    {\small $+(\mathbf{1},\mathbf{2},\mathbf{1},\mathbf{2},\mathbf{8}_v)+(\mathbf{1},\mathbf{2},\mathbf{1},\mathbf{2},\mathbf{8}_s)$}\\
    {\small$+(\mathbf{1},\mathbf{2},\mathbf{1},\mathbf{2},\mathbf{8}_c)$}\\ [+0.9mm]
\end{tabular} \\ \hline
& & \\ 
Spectrum & $G+9T+12V+24H_c$    &  $G+T+300V+544H_c$
\\[+1.7mm]\hline 
& & \\ 
Kodaira Condition & Yes & No \\ \hline
& & \\ 
\begin{tabular}{c}
Spectrum\\
after Higgsing 
\end{tabular}
 & 
 \begin{tabular}{c}
     $G+9T+8V+20H_0$  \\
     $U(1)^8$ gauge group 
\end{tabular}  
 & $G+T+244H_0$ 
\\[+0.7mm]\hline
& & \\ 
\begin{tabular}{c}
(would be)\\
F-dual\\
after Higgsing
\end{tabular}
& 
\begin{tabular}{c}
     {\small$h^{1,1}=h^{2,1}=19$,}\\ [+0.7mm]
     {\small elliptically fibered  $dP_9$ ( $\dfrac{(T^2)^3}{\mathbb{Z}_2^2}$)}\\ [+3mm]
\end{tabular}
& 
\begin{tabular}{c}
     {\small $h^{1,1}=3, h^{2,1}=243$,}\\ [+0.7mm]
     elliptically 
     fibered $\mathbb{F}_0 $ \footnotemark\\ [+0.7mm]
\end{tabular}
\\[+0.7mm]\hline
& & \\ 
Instanton number & N/A & $(12,12)$
\\[+1.3mm]\hline
\end{longtable}
\newpage

\begin{longtable}{|c|c|c|c|c|}
\hline
\begin{tabular}{c}
     Without  \\
     neutral $H_0$ 
\end{tabular}

& \cellcolor[HTML]{FFFFC7}  \textbf{Model} $3$ & \cellcolor[HTML]{FFFFC7}  \textbf{Model} $4$ 
\\ [+0.5mm]\hline 
Type &  Heterotic & Heterotic 
\\[+0.8mm]\hline
& & \\ 
Lattice & $\Gamma^{4,4}(A_2\oplus A_2)+2\Gamma^{8,0}(E_8)$ 
& $\Gamma^{4,4}(D_4)+2\Gamma^{8,0}(E_8)$
\\[+1.3mm]\hline
Twist &
\begin{tabular}{c}
     \\[-4mm]
     $\phi_L=(0,0),$  \\[+0.7mm]
     $\phi_R=(\frac{2}{3},\frac{2}{3})$  \\[+0.7mm]
\end{tabular}
&
\begin{tabular}{c}
     \\[-4mm]
     $\phi_L=(0,0),$  \\[+0.7mm]
     $E_8\leftrightarrow E_8,$  \\[+0.7mm]
     $\phi_R=(\frac{1}{2},\frac{1}{2})$  \\[+0.7mm]
\end{tabular}
\\[+0.7mm]\hline
& & \\ 
Shift & $V_L=\frac{1}{3}(1^6,0^2;0^8)$ & No shift
\\[+1.7mm]\hline 
& & \\ 
Gauge & {\small$E_6\times SU(3)\times E_8 \times SU(3)^2
$} & $E_8\times SO(8)$
\\[+1.3mm]\hline
\begin{tabular}{c}
Representation  \\

\end{tabular}
& 
\begin{tabular}{c}
     \\[-4mm]
     {\small $2(\mathbf{27},\mathbf{3},\mathbf{1},\mathbf{1},\mathbf{1})+(\mathbf{27},\mathbf{1},\mathbf{1},\underline{\mathbf{3},\mathbf{1}})$}\\[+0.7mm]
    {\small $+(\mathbf{27},\mathbf{1},\mathbf{1},\underline{\overline{\mathbf{3}},\mathbf{1}})+(\mathbf{1},\mathbf{3},\mathbf{1},\mathbf{3},\mathbf{3})$}\\[+0.7mm]
    {\small$+(\mathbf{1},\mathbf{3},\mathbf{1},\underline{\mathbf{3},\overline{\mathbf{3}}})+(\mathbf{1},\mathbf{3},\mathbf{1},\overline{\mathbf{3}},\overline{\mathbf{3}})$}\\[+0.7mm]
\end{tabular}
&
\begin{tabular}{c}
     \\[-4mm]
    {\small $2(\mathbf{248},\mathbf{1})+(\mathbf{1},\mathbf{8}_v)$}\\[+0.7mm]
    {\small $(\mathbf{1},\mathbf{8}_s)+(\mathbf{1},\mathbf{8}_c)$}\\[+0.7mm]
\end{tabular} \\ \hline
& & \\ 

Spectrum 
&  $G+T+350V+594H_c$ & $G+T+276V+520H_c$
\\[+1.7mm]\hline 
& & \\ 
Kodaira Condition & No & No \\ [+1.7mm ]\hline
& & \\ 
\begin{tabular}{c}
Spectrum\\
after Higgsing
\end{tabular}
& 
\begin{tabular}{c}
     $G+T+248V+492H_0$  \\
     $E_8$ gauge group
\end{tabular} 
& 
\begin{tabular}{c}
     $G+T+8V+252H_0$  \\
     $SU(3)$ gauge group
\end{tabular} 
\\[+1.7mm]\hline
& & \\ 
\begin{tabular}{c}
(would be)\\
F-dual\\
after Higgsing
\end{tabular}
& 
\begin{tabular}{c}
     $h^{1,1}=11, h^{2,1}=491$,\\ [+0.7mm]
     elliptically fibered $\mathbb{F}_{12}$
\end{tabular}
&
\begin{tabular}{c}
     $h^{1,1}=5, h^{2,1}= 251$,\\ [+0.7mm]
     elliptically fibered $\mathbb{F}_{3}$\\ [+0.7mm]
     with $(-3)$-curve
\end{tabular}
\\[+1.7mm]\hline 
& & \\ 
Instanton number & $(0,24)$ & $(9,15)$
\\[+1.3mm]\hline
 \caption{6d $\mathcal{N}=(1,0)$ models without neutral hypermultiplets. Here $\pm$ means charge $\pm 1/2$.  The hypermultiplets are all charged and their representation is listed in the table. The details of these models can be found in \autoref{app:AOdetails}.  The notation $G,V,T,H_0,H_c$ refers to gravity multiplet, vectors, tensors, neutral hypers, and charged hypers respectively. The details of the Higgsing process are described in \autoref{ap:maxhiggsing}.
The instanton number denotes the instanton numbers of the two $E_8$'s in the $E_8\times E_8$ heterotic model on K3.
}\label{tab:6d_modelA}
\end{longtable}
\renewcommand{\baselinestretch}{1}
 \footnotetext{Note that all $\mathbb{F}_n$ with $n=0,1,2$ give the same Hodge numbers for the threefold. However, the $\mathbb{Z}_2$ symmetry of the gauge group seems to indicate that both instanton numbers should be the same (based on how they embed on the two heterotic $E_8$'s )and hence needs  to be $n=0$. Note that according to \cite{Morrison:1996na} the $n=0$ and $n=2$ cases may represent the same physics. }

However, if one turns on  vevs for some of the charged hypermultiplets then the full gauge group can be Higgsed to a subgroup and some neutral hypermultiplets will remain which could potentially correspond to dual geometric models.  So in this way we connect a non-geometric model to a geometric model via Higgsing/unHiggsing.
Indeed examples of this type have been found in \cite{Kachru:1995wm} and here we find many more such examples as well as connect to various other dual descriptions.

The maximally Higgsed phase of these theories is described in detail \autoref{ap:maxhiggsing} and summarized in \autoref{tab:6d_modelA}. Then the number of multiplets can be matched with the expected Hodge numbers if the theory had a geometric description ala F-theory. Furthermore, Models 3 and 4 contain non-Higgsable clusters and hence the exact identification of the base of the elliptic threefold can be deduced.

 For example, Model 3 in the maximally Higgsed phase has a leftover $E_8$ gauge symmetry and no charged matter but $492$ free hypermultiplets which matches the expectation of the matter in the maximally Higgsed phase of F-theory on an elliptic threefold with base $\mathbb{F}_{12}$. 

The models 2, 3 and 4 which in the Higgsed phase correspond to F-theory on elliptic 3-fold with base $F_n$ are dual to $E_8\times E_8$ heterotic strings on $K3$ with instanton numbers $(12-n,12+n)$ in the two $E_8$'s.  Indeed for these models a distinct dual heterotic perspective offers another way to understand the hypermultiplets becoming charged at special points in the moduli (as was observed in \cite{Kachru:1995wm} for the specific case of model 3 in its 4d compactification):  It is the familiar story that as we vary the Narain moduli of heterotic strings, one gets enhanced gauge symmetries.  The Narain moduli correspond to hypermultiplets, including the universal one.  Thus the transition from geometric to the non-geometric phase is more easily understood from this perspective.  

 On the other hand, the first model has $T=9$ and hence does not come from the perturbative heterotic string. However, there is a fiber-wise duality to the heterotic string. In fact the base $dP_9$ corresponds to $\frac{1}{2}K3$ and is an elliptic surface with an infinite number of rational $(-1)$ curves which when combined give rise to $\mathbb {P}_1$'s which become the base of some $K3$ fibration. In fact as studied in \cite{Aspinwall:1996mw} there are infinitely many such $K3$ fibrations corresponding to different heterotic string limits. The heterotic limits are in fact equivalent through some U-duality group since the $K3$ fibrations are all diffeomorphic.
It would be interesting to know if the enhancement points could be understood from the heterotic picture (together with small instantons), at enhanced gauge symmetries as in Narain lattice compactifications.

\subsection{Compactifying from 6d down to 5d and 4d}
The 6d models become even more interesting when compactified on a circle or a torus.  In fact consider Models 1,2,3 on a circle, now one is allowed to turn on Wilson lines for  the gauge fields\footnote{The 6d vector multiplets have no scalars but the 5d $\mathcal{N}=1$ theory has vector multiplets with a real scalar.} and hence at  a generic point of the Coulomb branch the charged hypermultiplets acquire masses, and the gauge group is broken to the multiple $U(1)$s. In this way, we obtain the 5d theories without hypermultiplets starting from 6d theories with charged hypermultiplets. In fact the first three models all have rank 22 and no hypermultiplets which leads one to believe that they could all be describing the same theory. 
As for Model 4, the rank of the 5d theory is $14$ also with no massless hypers.

We can also connect this to the perspective of M-theory compactification on Calabi-Yau threefold in 5d, using the duality between F-theory on a circle and M-theory.  For example model 3, corresponds to M-theory on an elliptic fibration over $\mathbb{F}_{12}$.  As we just argued, when we change the volume of the M-theory compactification at special small size, the theory enjoys an enhanced gauge symmetry of $E_6\times SU(3)^3\times E_8$, for which the volume modulus is charged.  We can  then go to the Columb branch of this new enhanced gauge group, which in turn higgses all the charged hypermultiplets, leading to a model with rank 22 and no hypers.  This has no geometric description in M-theory and should be viewed as a Planckian sized geometry for the M-theory compactification, at which point the classical geometry picture is no longer applicable.

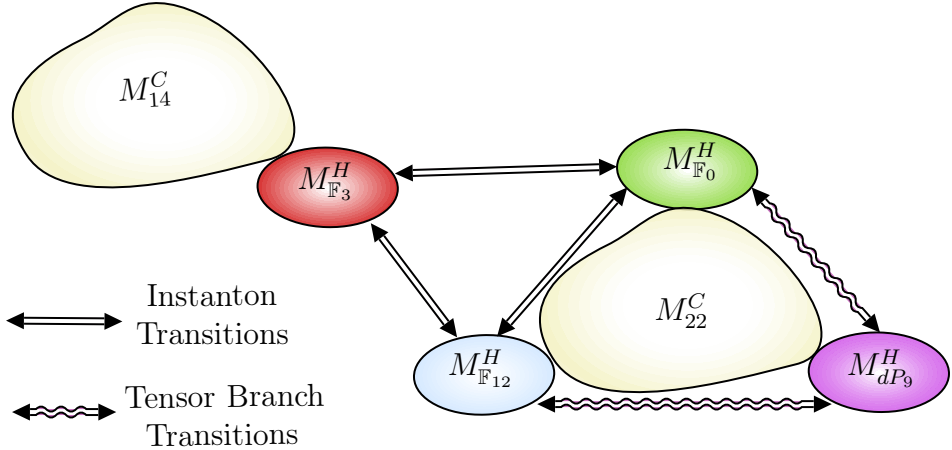
\begin{figure}
    \centering

  
\tikzset {_shiwntkkl/.code = {\pgfsetadditionalshadetransform{ \pgftransformshift{\pgfpoint{0 bp } { 0 bp }  }  \pgftransformscale{1 }  }}}
\pgfdeclareradialshading{_yuc4rkg1d}{\pgfpoint{0bp}{0bp}}{rgb(0bp)=(1,1,1);
rgb(0bp)=(1,1,1);
rgb(25bp)=(0.49,0.83,0.13);
rgb(400bp)=(0.49,0.83,0.13)}
\tikzset{_7nkkb5nmy/.code = {\pgfsetadditionalshadetransform{\pgftransformshift{\pgfpoint{0 bp } { 0 bp }  }  \pgftransformscale{1 } }}}
\pgfdeclareradialshading{_413cmz6n5} { \pgfpoint{0bp} {0bp}} {color(0bp)=(transparent!0);
color(0bp)=(transparent!0);
color(25bp)=(transparent!22);
color(400bp)=(transparent!22)} 
\pgfdeclarefading{_kh4960vs4}{\tikz \fill[shading=_413cmz6n5,_7nkkb5nmy] (0,0) rectangle (50bp,50bp); } 

  
\tikzset {_2ggpps7cw/.code = {\pgfsetadditionalshadetransform{ \pgftransformshift{\pgfpoint{0 bp } { 0 bp }  }  \pgftransformscale{1 }  }}}
\pgfdeclareradialshading{_7mx6c9w0w}{\pgfpoint{0bp}{0bp}}{rgb(0bp)=(1,1,1);
rgb(0bp)=(1,1,1);
rgb(25bp)=(0.74,0.06,0.88);
rgb(400bp)=(0.74,0.06,0.88)}
\tikzset{_yfqklca5s/.code = {\pgfsetadditionalshadetransform{\pgftransformshift{\pgfpoint{0 bp } { 0 bp }  }  \pgftransformscale{1 } }}}
\pgfdeclareradialshading{_7chkuk28z} { \pgfpoint{0bp} {0bp}} {color(0bp)=(transparent!0);
color(0bp)=(transparent!0);
color(25bp)=(transparent!40);
color(400bp)=(transparent!40)} 
\pgfdeclarefading{_fhw5i02gm}{\tikz \fill[shading=_7chkuk28z,_yfqklca5s] (0,0) rectangle (50bp,50bp); } 

  
\tikzset {_tynf0uxqf/.code = {\pgfsetadditionalshadetransform{ \pgftransformshift{\pgfpoint{0 bp } { 0 bp }  }  \pgftransformscale{1 }  }}}
\pgfdeclareradialshading{_0aeab17id}{\pgfpoint{0bp}{0bp}}{rgb(0bp)=(1,1,1);
rgb(0bp)=(1,1,1);
rgb(25bp)=(0.83,0.91,1);
rgb(400bp)=(0.83,0.91,1)}

  
\tikzset {_cniz3h8gx/.code = {\pgfsetadditionalshadetransform{ \pgftransformshift{\pgfpoint{0 bp } { 0 bp }  }  \pgftransformscale{1 }  }}}
\pgfdeclareradialshading{_3lmqxwkwu}{\pgfpoint{0bp}{0bp}}{rgb(0bp)=(1,1,1);
rgb(0bp)=(1,1,1);
rgb(25bp)=(0.84,0.17,0.17);
rgb(400bp)=(0.84,0.17,0.17)}

  
\tikzset {_o0cndtogx/.code = {\pgfsetadditionalshadetransform{ \pgftransformshift{\pgfpoint{0 bp } { 0 bp }  }  \pgftransformscale{1 }  }}}
\pgfdeclareradialshading{_cd4xlhd9f}{\pgfpoint{0bp}{0bp}}{rgb(0bp)=(1,1,1);
rgb(8.928571428571429bp)=(1,1,1);
rgb(20.089285714285715bp)=(0.95,0.94,0.73);
rgb(25bp)=(0.95,0.94,0.73);
rgb(400bp)=(0.95,0.94,0.73)}
\tikzset{_vn6m2olr8/.code = {\pgfsetadditionalshadetransform{\pgftransformshift{\pgfpoint{0 bp } { 0 bp }  }  \pgftransformscale{1 } }}}
\pgfdeclareradialshading{_qsngfxd14} { \pgfpoint{0bp} {0bp}} {color(0bp)=(transparent!48);
color(8.928571428571429bp)=(transparent!48);
color(20.089285714285715bp)=(transparent!23);
color(25bp)=(transparent!48);
color(400bp)=(transparent!48)} 
\pgfdeclarefading{_kqaue6xrp}{\tikz \fill[shading=_qsngfxd14,_vn6m2olr8] (0,0) rectangle (50bp,50bp); } 

  
\tikzset {_a7fmrlpex/.code = {\pgfsetadditionalshadetransform{ \pgftransformshift{\pgfpoint{0 bp } { 0 bp }  }  \pgftransformscale{1 }  }}}
\pgfdeclareradialshading{_vedq8ycz1}{\pgfpoint{0bp}{0bp}}{rgb(0bp)=(1,1,1);
rgb(8.928571428571429bp)=(1,1,1);
rgb(20.089285714285715bp)=(0.95,0.94,0.73);
rgb(25bp)=(0.95,0.94,0.73);
rgb(400bp)=(0.95,0.94,0.73)}
\tikzset{_1lpfsh49h/.code = {\pgfsetadditionalshadetransform{\pgftransformshift{\pgfpoint{0 bp } { 0 bp }  }  \pgftransformscale{1 } }}}
\pgfdeclareradialshading{_q0b3px28v} { \pgfpoint{0bp} {0bp}} {color(0bp)=(transparent!48);
color(8.928571428571429bp)=(transparent!48);
color(20.089285714285715bp)=(transparent!23);
color(25bp)=(transparent!48);
color(400bp)=(transparent!48)} 
\pgfdeclarefading{_24m5z29hq}{\tikz \fill[shading=_q0b3px28v,_1lpfsh49h] (0,0) rectangle (50bp,50bp); } 
\tikzset{every picture/.style={line width=0.75pt}} 

\begin{tikzpicture}[x=0.75pt,y=0.75pt,yscale=-1,xscale=1]

\path  [shading=_yuc4rkg1d,_shiwntkkl,path fading= _kh4960vs4 ,fading transform={xshift=2}] (317.87,200.55) .. controls (318.34,189.51) and (334.37,181.22) .. (353.68,182.04) .. controls (372.99,182.85) and (388.27,192.45) .. (387.81,203.49) .. controls (387.35,214.53) and (371.31,222.81) .. (352,222) .. controls (332.69,221.19) and (317.41,211.58) .. (317.87,200.55) -- cycle ; 
 \draw   (317.87,200.55) .. controls (318.34,189.51) and (334.37,181.22) .. (353.68,182.04) .. controls (372.99,182.85) and (388.27,192.45) .. (387.81,203.49) .. controls (387.35,214.53) and (371.31,222.81) .. (352,222) .. controls (332.69,221.19) and (317.41,211.58) .. (317.87,200.55) -- cycle ; 

\path  [shading=_7mx6c9w0w,_2ggpps7cw,path fading= _fhw5i02gm ,fading transform={xshift=2}] (413.5,303) .. controls (413.81,291.96) and (429.72,283.44) .. (449.04,283.98) .. controls (468.36,284.52) and (483.78,293.9) .. (483.47,304.94) .. controls (483.17,315.98) and (467.25,324.5) .. (447.93,323.96) .. controls (428.61,323.43) and (413.19,314.04) .. (413.5,303) -- cycle ; 
 \draw  [color={rgb, 255:red, 0; green, 0; blue, 0 }  ,draw opacity=1 ] (413.5,303) .. controls (413.81,291.96) and (429.72,283.44) .. (449.04,283.98) .. controls (468.36,284.52) and (483.78,293.9) .. (483.47,304.94) .. controls (483.17,315.98) and (467.25,324.5) .. (447.93,323.96) .. controls (428.61,323.43) and (413.19,314.04) .. (413.5,303) -- cycle ; 

\path  [shading=_0aeab17id,_tynf0uxqf] (286.5,304) .. controls (286.91,315.04) and (271.59,324.57) .. (252.27,325.29) .. controls (232.95,326.01) and (216.96,317.65) .. (216.55,306.61) .. controls (216.14,295.58) and (231.46,286.04) .. (250.78,285.32) .. controls (270.09,284.6) and (286.09,292.96) .. (286.5,304) -- cycle ; 
 \draw   (286.5,304) .. controls (286.91,315.04) and (271.59,324.57) .. (252.27,325.29) .. controls (232.95,326.01) and (216.96,317.65) .. (216.55,306.61) .. controls (216.14,295.58) and (231.46,286.04) .. (250.78,285.32) .. controls (270.09,284.6) and (286.09,292.96) .. (286.5,304) -- cycle ; 

\draw [fill={rgb, 255:red, 189; green, 16; blue, 224 }  ,fill opacity=1 ]   (318.24,218.78) -- (265.03,280.18)(315.97,216.82) -- (262.76,278.22) ;
\draw [shift={(258,286)}, rotate = 310.91] [fill={rgb, 255:red, 0; green, 0; blue, 0 }  ][line width=0.08]  [draw opacity=0] (8.93,-4.29) -- (0,0) -- (8.93,4.29) -- cycle    ;
\draw [shift={(323,211)}, rotate = 130.91] [fill={rgb, 255:red, 0; green, 0; blue, 0 }  ][line width=0.08]  [draw opacity=0] (8.93,-4.29) -- (0,0) -- (8.93,4.29) -- cycle    ;
\draw [fill={rgb, 255:red, 189; green, 16; blue, 224 }  ,fill opacity=1 ]   (392.06,216.78) -- (394.04,219.03) .. controls (396.39,219.18) and (397.49,220.44) .. (397.34,222.79) .. controls (397.19,225.14) and (398.29,226.4) .. (400.64,226.55) .. controls (402.99,226.7) and (404.09,227.95) .. (403.94,230.3) .. controls (403.78,232.65) and (404.88,233.91) .. (407.23,234.06) .. controls (409.58,234.21) and (410.68,235.47) .. (410.53,237.82) .. controls (410.38,240.17) and (411.48,241.43) .. (413.83,241.58) .. controls (416.18,241.73) and (417.28,242.99) .. (417.13,245.34) .. controls (416.98,247.69) and (418.08,248.95) .. (420.43,249.1) .. controls (422.78,249.25) and (423.87,250.5) .. (423.72,252.85) .. controls (423.57,255.2) and (424.67,256.46) .. (427.02,256.61) .. controls (429.37,256.76) and (430.47,258.02) .. (430.32,260.37) .. controls (430.17,262.72) and (431.27,263.98) .. (433.62,264.13) .. controls (435.97,264.28) and (437.07,265.54) .. (436.92,267.89) .. controls (436.77,270.24) and (437.86,271.49) .. (440.21,271.64) -- (442.25,273.97) -- (444.23,276.22)(389.81,218.75) -- (391.79,221.01) .. controls (394.14,221.16) and (395.24,222.42) .. (395.09,224.77) .. controls (394.93,227.12) and (396.03,228.38) .. (398.38,228.53) .. controls (400.73,228.68) and (401.83,229.93) .. (401.68,232.28) .. controls (401.53,234.63) and (402.63,235.89) .. (404.98,236.04) .. controls (407.33,236.19) and (408.43,237.45) .. (408.28,239.8) .. controls (408.13,242.15) and (409.23,243.41) .. (411.58,243.56) .. controls (413.93,243.71) and (415.03,244.97) .. (414.87,247.32) .. controls (414.72,249.67) and (415.82,250.92) .. (418.17,251.07) .. controls (420.52,251.22) and (421.62,252.48) .. (421.47,254.83) .. controls (421.32,257.18) and (422.42,258.44) .. (424.77,258.59) .. controls (427.12,258.74) and (428.22,260) .. (428.06,262.35) .. controls (427.91,264.7) and (429.01,265.96) .. (431.36,266.11) .. controls (433.71,266.26) and (434.81,267.52) .. (434.66,269.87) .. controls (434.51,272.22) and (435.61,273.47) .. (437.96,273.62) -- (440,275.95) -- (441.98,278.2) ;
\draw [shift={(449.04,283.98)}, rotate = 228.73] [fill={rgb, 255:red, 0; green, 0; blue, 0 }  ][line width=0.08]  [draw opacity=0] (8.93,-4.29) -- (0,0) -- (8.93,4.29) -- cycle    ;
\draw [shift={(385,211)}, rotate = 48.73] [fill={rgb, 255:red, 0; green, 0; blue, 0 }  ][line width=0.08]  [draw opacity=0] (8.93,-4.29) -- (0,0) -- (8.93,4.29) -- cycle    ;
\draw [fill={rgb, 255:red, 189; green, 16; blue, 224 }  ,fill opacity=1 ]   (287,318.5) -- (290,318.5) .. controls (291.67,316.83) and (293.33,316.83) .. (295,318.5) .. controls (296.67,320.17) and (298.33,320.17) .. (300,318.5) .. controls (301.67,316.83) and (303.33,316.83) .. (305,318.5) .. controls (306.67,320.17) and (308.33,320.17) .. (310,318.5) .. controls (311.67,316.83) and (313.33,316.83) .. (315,318.5) .. controls (316.67,320.17) and (318.33,320.17) .. (320,318.5) .. controls (321.67,316.83) and (323.33,316.83) .. (325,318.5) .. controls (326.67,320.17) and (328.33,320.17) .. (330,318.5) .. controls (331.67,316.83) and (333.33,316.83) .. (335,318.5) .. controls (336.67,320.17) and (338.33,320.17) .. (340,318.5) .. controls (341.67,316.83) and (343.33,316.83) .. (345,318.5) .. controls (346.67,320.17) and (348.33,320.17) .. (350,318.5) .. controls (351.67,316.83) and (353.33,316.83) .. (355,318.5) .. controls (356.67,320.17) and (358.33,320.17) .. (360,318.5) .. controls (361.67,316.83) and (363.33,316.83) .. (365,318.5) .. controls (366.67,320.17) and (368.33,320.17) .. (370,318.5) .. controls (371.67,316.83) and (373.33,316.83) .. (375,318.5) .. controls (376.67,320.17) and (378.33,320.17) .. (380,318.5) .. controls (381.67,316.83) and (383.33,316.83) .. (385,318.5) .. controls (386.67,320.17) and (388.33,320.17) .. (390,318.5) .. controls (391.67,316.83) and (393.33,316.83) .. (395,318.5) .. controls (396.67,320.17) and (398.33,320.17) .. (400,318.5) .. controls (401.67,316.83) and (403.33,316.83) .. (405,318.5) .. controls (406.67,320.17) and (408.33,320.17) .. (410,318.5) -- (413,318.5) -- (416,318.5)(287,321.5) -- (290,321.5) .. controls (291.67,319.83) and (293.33,319.83) .. (295,321.5) .. controls (296.67,323.17) and (298.33,323.17) .. (300,321.5) .. controls (301.67,319.83) and (303.33,319.83) .. (305,321.5) .. controls (306.67,323.17) and (308.33,323.17) .. (310,321.5) .. controls (311.67,319.83) and (313.33,319.83) .. (315,321.5) .. controls (316.67,323.17) and (318.33,323.17) .. (320,321.5) .. controls (321.67,319.83) and (323.33,319.83) .. (325,321.5) .. controls (326.67,323.17) and (328.33,323.17) .. (330,321.5) .. controls (331.67,319.83) and (333.33,319.83) .. (335,321.5) .. controls (336.67,323.17) and (338.33,323.17) .. (340,321.5) .. controls (341.67,319.83) and (343.33,319.83) .. (345,321.5) .. controls (346.67,323.17) and (348.33,323.17) .. (350,321.5) .. controls (351.67,319.83) and (353.33,319.83) .. (355,321.5) .. controls (356.67,323.17) and (358.33,323.17) .. (360,321.5) .. controls (361.67,319.83) and (363.33,319.83) .. (365,321.5) .. controls (366.67,323.17) and (368.33,323.17) .. (370,321.5) .. controls (371.67,319.83) and (373.33,319.83) .. (375,321.5) .. controls (376.67,323.17) and (378.33,323.17) .. (380,321.5) .. controls (381.67,319.83) and (383.33,319.83) .. (385,321.5) .. controls (386.67,323.17) and (388.33,323.17) .. (390,321.5) .. controls (391.67,319.83) and (393.33,319.83) .. (395,321.5) .. controls (396.67,323.17) and (398.33,323.17) .. (400,321.5) .. controls (401.67,319.83) and (403.33,319.83) .. (405,321.5) .. controls (406.67,323.17) and (408.33,323.17) .. (410,321.5) -- (413,321.5) -- (416,321.5) ;
\draw [shift={(425,320)}, rotate = 180] [fill={rgb, 255:red, 0; green, 0; blue, 0 }  ][line width=0.08]  [draw opacity=0] (8.93,-4.29) -- (0,0) -- (8.93,4.29) -- cycle    ;
\draw [shift={(278,320)}, rotate = 0] [fill={rgb, 255:red, 0; green, 0; blue, 0 }  ][line width=0.08]  [draw opacity=0] (8.93,-4.29) -- (0,0) -- (8.93,4.29) -- cycle    ;
\path  [shading=_3lmqxwkwu,_cniz3h8gx] (138.5,210) .. controls (138.81,198.96) and (154.72,190.44) .. (174.04,190.98) .. controls (193.36,191.52) and (208.78,200.9) .. (208.47,211.94) .. controls (208.17,222.98) and (192.25,231.5) .. (172.93,230.96) .. controls (153.61,230.43) and (138.19,221.04) .. (138.5,210) -- cycle ; 
 \draw   (138.5,210) .. controls (138.81,198.96) and (154.72,190.44) .. (174.04,190.98) .. controls (193.36,191.52) and (208.78,200.9) .. (208.47,211.94) .. controls (208.17,222.98) and (192.25,231.5) .. (172.93,230.96) .. controls (153.61,230.43) and (138.19,221.04) .. (138.5,210) -- cycle ; 

\draw [fill={rgb, 255:red, 189; green, 16; blue, 224 }  ,fill opacity=1 ]   (201.6,235.3) -- (233.86,278.32)(199.2,237.1) -- (231.46,280.12) ;
\draw [shift={(238.06,286.42)}, rotate = 233.13] [fill={rgb, 255:red, 0; green, 0; blue, 0 }  ][line width=0.08]  [draw opacity=0] (8.93,-4.29) -- (0,0) -- (8.93,4.29) -- cycle    ;
\draw [shift={(195,229)}, rotate = 53.13] [fill={rgb, 255:red, 0; green, 0; blue, 0 }  ][line width=0.08]  [draw opacity=0] (8.93,-4.29) -- (0,0) -- (8.93,4.29) -- cycle    ;
\draw [fill={rgb, 255:red, 189; green, 16; blue, 224 }  ,fill opacity=1 ]   (308.92,202.33) -- (216.04,205.22)(308.83,199.33) -- (215.95,202.22) ;
\draw [shift={(207,204)}, rotate = 358.22] [fill={rgb, 255:red, 0; green, 0; blue, 0 }  ][line width=0.08]  [draw opacity=0] (8.93,-4.29) -- (0,0) -- (8.93,4.29) -- cycle    ;
\draw [shift={(317.87,200.55)}, rotate = 178.22] [fill={rgb, 255:red, 0; green, 0; blue, 0 }  ][line width=0.08]  [draw opacity=0] (8.93,-4.29) -- (0,0) -- (8.93,4.29) -- cycle    ;
\draw [fill={rgb, 255:red, 189; green, 16; blue, 224 }  ,fill opacity=1 ]   (22.01,276.56) -- (62.07,276.85)(21.99,279.56) -- (62.05,279.85) ;
\draw [shift={(71.06,278.42)}, rotate = 180.41] [fill={rgb, 255:red, 0; green, 0; blue, 0 }  ][line width=0.08]  [draw opacity=0] (8.93,-4.29) -- (0,0) -- (8.93,4.29) -- cycle    ;
\draw [shift={(13,278)}, rotate = 0.41] [fill={rgb, 255:red, 0; green, 0; blue, 0 }  ][line width=0.08]  [draw opacity=0] (8.93,-4.29) -- (0,0) -- (8.93,4.29) -- cycle    ;
\draw [fill={rgb, 255:red, 189; green, 16; blue, 224 }  ,fill opacity=1 ]   (25,322.5) -- (28,322.5) .. controls (29.67,320.83) and (31.33,320.83) .. (33,322.5) .. controls (34.67,324.17) and (36.33,324.17) .. (38,322.5) .. controls (39.67,320.83) and (41.33,320.83) .. (43,322.5) .. controls (44.67,324.17) and (46.33,324.17) .. (48,322.5) .. controls (49.67,320.83) and (51.33,320.83) .. (53,322.5) -- (56,322.5) -- (59,322.5)(25,325.5) -- (28,325.5) .. controls (29.67,323.83) and (31.33,323.83) .. (33,325.5) .. controls (34.67,327.17) and (36.33,327.17) .. (38,325.5) .. controls (39.67,323.83) and (41.33,323.83) .. (43,325.5) .. controls (44.67,327.17) and (46.33,327.17) .. (48,325.5) .. controls (49.67,323.83) and (51.33,323.83) .. (53,325.5) -- (56,325.5) -- (59,325.5) ;
\draw [shift={(68,324)}, rotate = 180] [fill={rgb, 255:red, 0; green, 0; blue, 0 }  ][line width=0.08]  [draw opacity=0] (8.93,-4.29) -- (0,0) -- (8.93,4.29) -- cycle    ;
\draw [shift={(16,324)}, rotate = 0] [fill={rgb, 255:red, 0; green, 0; blue, 0 }  ][line width=0.08]  [draw opacity=0] (8.93,-4.29) -- (0,0) -- (8.93,4.29) -- cycle    ;
\path  [shading=_cd4xlhd9f,_o0cndtogx,path fading= _kqaue6xrp ,fading transform={xshift=2}] (313,240) .. controls (333,230) and (341,212) .. (373,227) .. controls (405,242) and (439,292) .. (407,300) .. controls (375,308) and (306.5,328) .. (286.5,298) .. controls (266.5,268) and (293,250) .. (313,240) -- cycle ; 
 \draw   (313,240) .. controls (333,230) and (341,212) .. (373,227) .. controls (405,242) and (439,292) .. (407,300) .. controls (375,308) and (306.5,328) .. (286.5,298) .. controls (266.5,268) and (293,250) .. (313,240) -- cycle ; 

\path  [shading=_vedq8ycz1,_a7fmrlpex,path fading= _24m5z29hq ,fading transform={xshift=2}] (50,137) .. controls (70,127) and (78,109) .. (110,124) .. controls (142,139) and (176,189) .. (144,197) .. controls (112,205) and (43.5,225) .. (23.5,195) .. controls (3.5,165) and (30,147) .. (50,137) -- cycle ; 
 \draw   (50,137) .. controls (70,127) and (78,109) .. (110,124) .. controls (142,139) and (176,189) .. (144,197) .. controls (112,205) and (43.5,225) .. (23.5,195) .. controls (3.5,165) and (30,147) .. (50,137) -- cycle ; 

\draw (319,259) node [anchor=north west][inner sep=0.75pt]   [align=left] { };
\draw (339,187) node [anchor=north west][inner sep=0.75pt]    {$M_{\mathbb{F}_{0}}^{H}$};
\draw (431,291) node [anchor=north west][inner sep=0.75pt]    {$M_{dP_{9}}^{H}$};
\draw (232,291) node [anchor=north west][inner sep=0.75pt]    {$M_{\mathbb{F}_{12}}^{H}$};
\draw (76.32,258.01) node [anchor=north west][inner sep=0.75pt]  [rotate=-359.74] [align=left] { \ Instanton \\Transitions};
\draw (70.44,309.4) node [anchor=north west][inner sep=0.75pt]  [rotate=-359.24] [align=left] {Tensor Branch \\ \ \ Transitions};
\draw (157,197) node [anchor=north west][inner sep=0.75pt]    {$M_{\mathbb{F}_{3}}^{H}$};
\draw (335,263) node [anchor=north west][inner sep=0.75pt]    {$M_{22}^{C}$};
\draw (66.52,152.93) node [anchor=north west][inner sep=0.75pt]  [rotate=-356.63]  {$M_{14}^{C}$};

\end{tikzpicture}
    \caption{    This figure depicts the transitions between the geometric and non-geometric phases. The Higgs Branch of the geometric phases of the elliptic Calabi-Yau's with bases $\mathbb{F}_n$ or $dP_9$ from \autoref{tab:6d_modelA} are connected  to the asymmetric orbifolds  on their Coulomb branch when compactified on a circle with Wilson lines. The threefolds with bases $\mathbb{F}_n$ are connected through instanton transitions and the $dP_9$ threefold can be reduced to $\mathbb{F}_n$ through the tensor branch via successive blow downs and instanton transitions. The Calabi-Yau threefolds with base $\mathbb{F}_n$   have one tensor multiplet and hence a dual heterotic on K3 description with different configuration of instanton numbers $n_1,n_2$ e.g. $\mathbb{F}_{12}$ corresponds to $n_1=0,n_2=24$. The asymmetric orbifold can be reached directly from the heterotic discretion in these cases.}
    \label{fig:transitions}
\end{figure}

 A graphical depiction of these transitions is shown in \autoref{fig:transitions}. Moreover,  the last three F-theory models of \autoref{tab:6d_modelA} are connected through instanton transitions as described in \cite{Seiberg:1996vs,Ganor:1996mu,Morrison:1996pp} where one needs to go through some tensor branch transitions. Geometrically, this corresponds to a flop transition in the base where one needs to blow up  a curve and then blow down a different curve. Physically, at the origin of some Tensor Branch new Higgs branches appear and vice versa giving rise to such transitions.
 Additionally, the first model corresponding to an elliptic threefold with base $dP_9$ can be related to the $\mathbb{F}_n$ models through similar tensor branch transitions where one exchanges one tensor multiplet for some neutral hypermultiplets (charged hyper and vector could also appear). In geometric language these correspond to successive blow downs.

From the picture above, we are also able to figure out the strong coupling limit of the non-geometric models.  For example consider Model $2$ which corresponds to $\mathbb{F}_0=\mathbb{P}^1\times\mathbb{P}^1$. Since the tensor branch and Higgs branch do not talk to each other, the strong coupling limit before the Higgsing is the same as the one after the Higgsing. After the Higgsing, the model exhibits the S-duality exchanging the two $\mathbb{P}^1$s. Therefore, we conclude that the model is self-dual and that the strong coupling limit of our non-geometric models is the tensionless string limit.

Another interesting remark as described earlier in the section is that any F-theory compactification satisfies the Kodaira condition \autoref{eq:kodairaJ} but as it can be easily checked, \autoref{app:AOdetails}, the Models 2, 3 and 4 do not satisfy it. In fact the orbifold models tend to have a large amount of non-abelian enhancements which makes  all the hypers charged and hence the right hand side of \autoref{eq:kodaira} has the potential to become large violating the condition. These are good demonstrations of geometric consistency conditions being violated   in non-geometric theories.

\newpage
\section{5d models without hypermultiplets}\label{sec:5d}

In the previous section the Landscape of 6d $\mathcal{N}=(1,0)$ with no neutral hypermultiplets was explored using asymmetric orbifold techniques. Interestingly,   we saw that when compactified on a circle the non-geometric and geometric phases can be understood as part of a single moduli space.
In this section, we would like to extend our study to the Landscape of the 5d $\mathcal{N}=1$ supergravity theories. A large number of theories is realized through M-theory on Calabi-Yau threefolds. These theories contain the "universal hypermultiplet", as the number of hypermultiplets is given by $h^{2,1}+1$. However, we demonstrate that non-geometric asymmetric orbifolds yield models without hypermultiplets.  This provides further evidence that geometric compactification alone is not sufficient to capture all aspects of the QG landscape.  However, as discussed in the previous section  these non-geometric phases can be connected to geometric phases through strong coupling transitions involving Higgsing/unHiggsing.

We would like to mention that the 5d models without hypermultiplets were recently investigated in \cite{Gkountoumis:2023fym}, where  models with ranks 6, 8, and 14 were presented. However, we also study freely acting $\mathbb{Z}_N$ orbifold models and provide a more comprehensive list, as demostrated in Table~\ref{Tab:Freely_acting}.

An interesting observation is that the models with no neutral hypermultiplets seem to behave similar to their higher supersymmetric versions where the rank is bounded by $r_G\leq 26-d$ \cite{Kim:2019ths} which in our case takes the form $r_G\leq 26-d+1$ due to the extra dual tensor. This bound for the case of freely acting orbifolds can be understood as having a "memory" of the higher supersymmetric version of it similar to the discussions in \cite{Blumenhagen:2016rof}. However, in the previous section the 6d models that also gave maximal rank $22$ in 5d were not freely acting and hence the reasoning is not clear. It would be interesting to understand whether this bound is a universal feature of theories with no neutral hypers or an artifact of our constructions.
Additionally, as also observed in  \cite{Gkountoumis:2023fym} these hyper-free theories in 5d seem to allow only even ranks. Theories with  16 supercharges in 6d are suspected to also satisfy a similar condition.

If this is true and freely acting orbifolds do behave as their higher supersymmetric versions then it is natural to suspect that in fact this parity condition is universal also for hyper-free models in 5d.

\subsection*{5d freely acting  $\mathbb{Z}_N$ orbifolds  with no hypermultiplets.}

The asymmetric orbifold constructions we consider are similar to the previous section but now we focus on freely acting versions. In particular, we consider a 4d reduction with an extra circle on which the abelian discrete  $\mathbb{Z}_N$ twist acts as a shift which makes the  twisted sector  massive for a sufficiently large radius (see Appendix~\ref{sec:shift}).

In the case of type II models, we consider the momentum lattice
\begin{align}
    \Gamma^{5,5}=\Gamma^{4,4}+\Gamma^{1,1}.
\end{align}
and for the heterotic models, we use 
\begin{align}
    \Gamma^{21,5}=\Gamma^{20,4}+\Gamma^{1,1}.
\end{align}
The $\mathbb{Z}_N$ acts as a twist for $\Gamma^{4,4}$ and $\Gamma^{20,4}$, and acts as a shift for $\Gamma^{1,1}$ and hence lifting the twisted sectors.

The goal is to find a set of 5d models with 8 supercharges and no hypermultiplets. 
The details of the constructions are described in \autoref{ap:5dmodels} and summarized in \autoref{Tab:Freely_acting}.
The columns of the table represent the rank of the 5d theory, whether it descends from type II or heterotic string theory, the choice of momentum lattice and the  orbifold $\mathbb{Z}_N$ twist and shift. In contrast to the previous section since there are no invariant lattices, no shifts are present in the right mover while the shift in $\Gamma^{1,1}$ is implicitly understood.

 \renewcommand{\baselinestretch}{1.5}
\begin{longtable}{|c|c|c|c|c|c|}
\hline
rank &  type & 
\begin{tabular}{c}
     lattice \\
     $+\Gamma^{1,1}$
\end{tabular} 
&  twist & 
order & Ref.
\\
\hline

\hline

$2$ & II & $\Gamma^{4,4}(D_4)$ & 
\begin{tabular}{c}
     \\[-4mm]
     $\phi_L=(\frac{1}{6},\frac{3}{6})$  \\
     [+0.5mm]
     $\phi_R=(\frac{1}{4},\frac{1}{4})$ \\[+0.5mm] 
\end{tabular}
& $12$ & \\
[+0.3mm]
\hline
$4$ & II & $\Gamma^{4,4}(D_4)$ 
& 
\begin{tabular}{c}
     \\[-4mm]
     $\phi_L=(0,\frac{2}{3})$  \\
     [+0.7mm]
     $\phi_R=(\frac{1}{4},\frac{1}{4})$ \\[+0.5mm]
\end{tabular}
& $12$ &\\
[+0.3mm]\hline
$6$ & II & $\Gamma^{4,4}(A_2\times A_2)$ &
\begin{tabular}{c}
     \\[-4mm]
     $\phi_L=(1,0)$  \\
     [+0.7mm]
     $\phi_R=(\frac{1}{3},\frac{1}{3})$ \\[+0.5mm]
\end{tabular}
& $6$ & \cite{Gkountoumis:2023fym}
\\
[+0.3mm]\hline
$8$ & II & $\Gamma^{4,4}(D_4)$ & 
\begin{tabular}{c}
     \\[-4mm]
     $\phi_L=(\frac{1}{2},0)$  \\
     [+0.7mm]
     $\phi_R=(\frac{1}{4},\frac{1}{4})$ \\[+0.5mm]
\end{tabular}
& $4$ & \cite{Gkountoumis:2023fym}
\\[+0.3mm]\hline
\begin{tabular}{c}
     $12$ 
\end{tabular}
& Het & $2\Gamma^{2,2}(A_2)+2\Gamma^{8,0}(E_8)$
&
\begin{tabular}{c}
     \\[-4mm]
     $\phi_R=\left(\frac{1}{6},\frac{1}{6}\right)$  \\[+0.7mm] 
     $\Gamma^{2,2}(A_2)\leftrightarrow\Gamma^{2,2}(A_2)$\\[+0.7mm]
     $\Gamma^{8,0}(E_8)\leftrightarrow\Gamma^{8,0}(E_8)$ \\[+0.7mm]
     $V_L=(0^8;0^8)$\\[+0.5mm]
\end{tabular}
&  $6$ &
\\[+0.3mm]\hline
\multirow{3}{*}{$14$}
& Het & 
\begin{tabular}{c}
     \\[-4mm]
     $\Gamma^{4,4}(D_4)+2\Gamma^{8,0}(E_8)$    \\[+0.5mm]
\end{tabular}
&  
\begin{tabular}{c}
     \\[-4mm]
     $\phi_L=(0,0)$  \\
     [+0.7mm]
     $\phi_R=(\frac{1}{4},\frac{1}{4})$ \\[+0.7mm]
     $\Gamma^{8,0}(E_8)\leftrightarrow\Gamma^{8,0}(E_8)$\\[+0.7mm]
     $V_L=(0^8;0^8)$\\[+0.5mm]
\end{tabular}
& $4$ &
\\ [+1mm]\cline{2-5}
& II & 
\begin{tabular}{c}
Model $4$ on $S^1$, \\[+0.7mm]
Coulomb branch
\end{tabular}
& & &
\\[+0.3mm]\hline
$20$ & Het & $2\Gamma^{2,2}(A_2)+2\Gamma^{8,0}(E_8)$
&
\begin{tabular}{c}
     \\[-4mm]
     $\phi_R=(\frac{1}{6},\frac{1}{6})$  \\[+0.7mm] 
     $\Gamma^{2,2}(A_2)\leftrightarrow\Gamma^{2,2}(A_2)$\\[+0.7mm]
     $V_L=(0^8;0^8)$\\[+0.5mm]
\end{tabular}
&  $6$ &
\\[+0.3mm]\hline
\multirow{3}{*}{$22$}
& Het & $\Gamma^{4,4}(D_4)+2\Gamma^{8,0}(E_8)$ 
&
\begin{tabular}{c}
     \\[-4mm]
     $\phi_L=(0,0)$  \\[+0.7mm]
     $\phi_R=(\frac{1}{2},\frac{1}{2})$ \\[+0.7mm]
     $V_L=(0^8;0^8)$\\[+0.5mm]
\end{tabular}
& $2$ &
\\ [+1mm]\cline{2-5}
& II & 
\begin{tabular}{c}
Model $1,2,3$ on $S^1$, \\[+0.7mm]
Coulomb branch
\end{tabular}
& & &\\[+1mm]\hline
\caption{5d freely acting orbifold models without hypermultiplets. The models with rank $6$ and $8$ are the ones in \cite{Gkountoumis:2023fym} (Another realization of rank $14$ model is also provided there). The other models are new. It should be noted that the choice of the orbifold is not unique for each rank. The compactification of the models on a circle results in 4d models with odd ranks, which do not have hypermultiplets.}
\label{Tab:Freely_acting}
\end{longtable}
\renewcommand{\baselinestretch}{1}

As we noted before, only even rank theories (from $2$ to $22$) are obtained for 5d. 
Notice that the models with rank 6 ($\mathbb{Z}_6$ orbifold), 8 ($\mathbb{Z}_4$ orbifold) and a different rank 14 ($\mathbb{Z}_2$ orbifold) are given in \cite{Gkountoumis:2023fym}.

An interesting observation is that the minimal rank 2 example reduces to rank 3 in 4d with again no hypermultiplets making all the possible combinations $r+H=3$ realizable.  In fact this case was already realized in \cite{Israel:2013wwa} by the STU example which could potentially describe the same theory.  The other combinations can be understood through Calabi-Yau manifolds \cite{Candelas:2016fdy} with $(h^{1,1},h^{1,2})=(1,1),(2,0)$ corresponding to $(r,H)=(1,2),(2,1)$ and $(r,H)=(2,1),(1,2)$ respectively with the two choices describing whether we are compactifying IIA or IIB on the corresponding manifold.
It would be interesting to understand if examples with $r+H<3$ are possible. But we leave this discussion for the next section where 4d models will be analyzed in more detail.

It would be tempting to explain the parity of the rank of the vector multiplets in theories with no hypers using an anomaly.  Note that when we have one hypermultiplet, we have examples with both odd and even numbers of vector multiplets. Indeed \cite{Candelas:2016fdy} provides CY threefolds with $h^{1,1}=13, h^{2,1}=0$ and $h^{1,1}=2, h^{2,1}=0$, where M-theory would lead to even and odd vector rank vector multiplets with one hypermultiplet.  Therefore this implies that there cannot be a simple anomaly argument, in the presence of hypermultiplets.  The only option would be a symmetry that exists if and only if there are no hypermultiplets, and we are not aware of any such symmetry.

It is also curious that we are having the same bound on the rank as the theory with higher supersymmetry.  In our heterotic constructions, the freely acting orbifolds inherit the behavior of their higher supersymmetric versions. This could also explain why in these cases with no hypers the rank is bounded by $r_G\leq 26-d+1$ as the equivalent $r_G\leq 26-d$ bound holds for theories with 16 supercharges as shown in \cite{Kim:2019ths} with the extra vector coming from the additional tensor dual. But this condition would still only hold for freely acting cases and would not encompass examples like those given in the previous section which are not freely acting. Therefore, it would be interesting to either construct cases that violate these bounds or explain why the evenness of rank as well as the bound on the rank of the vector multiplet is a universal feature of theories with no hypers in 5d ${\cal N}=1$ in the context of the Swampland program.

\section{4d models without hypermultiplets}\label{sec:4d}
In this Section, we examine asymmetric orbifold constructions of 4d $\mathcal{N}=2$ theories without hypermultiplets. We investigate three different model types. 

\subsection{\texorpdfstring{$S^1$}{S1} compactification of 5d models}
An obvious construction of 4d hyper-free models is the $S^1$ compactification of 5d hyper-free models in Table~\ref{Tab:Freely_acting}. The $S^1$ compactification adds one vector multiplet. Therefore, we obtain 4d hyper-free models with odd number of vector multiplets, ranging in rank from $3$ to $23$.
Among these models, the rank $23$ model is previously constructed in \cite{Kachru:1995wm}.

\subsection{Intrinsically 4d model}
As just noted the circle compactifications of the 5d models we have constructed with no hypers yield a minimal number of vectors, namely 3.  A natural question is if we can achieve the rank$<3$ by considering an intrinsically 4d model. In particular, a no hyper model with one vector multiplet is interesting, since it can potentially lead to the 5d pure supergravity at the strong coupling limit~\cite{Dabholkar:1998kv}.
Here we demonstrate the absence of one-vector models within cyclic asymmetric orbifolds.

In order to realize 4d $\mathcal{N}=2$ model, we should preserve eight supercharges in the right-moving (or left-moving) sector.\footnote{Models with four supercharges in both left and right-moving sectors are likely to lead to hypermultplets.} This requirement indicates that the right-moving twist involves two eigenvalues of one, leading to two vector fields in the untwisted sector. One of them is the graviphoton, and the other is part of the vector multiplet. Consequently, in order to realize our goal, we need to perform left-moving twists without an invariant lattice. Otherwise, there will be additional vector fields in the untwisted sector. Furthermore, the addition of a shift vector is necessary to lift all the twisted sector states.

In the case of type II, this possibility is examined in \cite{Mizoguchi:2001cp}.
The author begins by enumerating all 4D $\mathcal{N}=4$ pure supergravity models derived from cyclic asymmetric orbifolds, then investigate whether further twist/shift leads to 4d $\mathcal{N}=2$ supergravity with one vector multiplet. The author finds no such twist/shift. 

It is instructive to see an example to get an intuition why it is difficult to realize one vector model. It is known that the 6d $\mathcal{N}=(1,1)$ pure supergravity is constructed as type II $\mathbb{Z}_5$ asymmetric orbifold with $\Gamma^{4,4}(A_4)$ lattice and the twist $\phi_L=(2/5,4/5)$ and $\phi_R=(0,0)$~\cite{Dabholkar:1998kv}.
One may think that the reduction of this model leads to the 4d $\mathcal{N}=2$ with one vector multiple. However, it turns out that this cannot be the case. If we choose $\Gamma^{6,6}(A_4\oplus A_2)$ lattice, then the left-moving twist must be
\begin{align}
    &\phi_L=\left(\frac{2}{5},\frac{4}{5},\frac{2}{3}\right),
\end{align}
to avoid the additional vector fields. Moreover, the right-moving twist should be
\begin{align}
    &\phi_R=\left(0,\frac{2}{3},\frac{2}{3}\right),
    &&\text{or}
    &&\left(0,\frac{1}{2},\frac{1}{2}\right)
\end{align}
to preserve the eight numbers of supersymmetry.
However both choices lead to extra fields in the untwisted/twisted sectors.
In Appendix~\ref{sec:no_4d_model}, we show that there are no models with one vector multiplet in the models based on the cyclic asymmetric orbifolds for type II or heterotic strings.  Moreover the rank of these models for cyclic orbifold models is always odd\footnote{This is because nonzero eigenvalues appear in complex pairs.}.  It would be interesting to see if this is a general feature or not.

\subsection{Speculative 4d models as a strong coupling limit of 3d models}\label{sec:3d_models}
Up to this point, we have not encountered any 4D no-hyper models with a rank smaller than $3$. Here we investigate the possibility of decompactifying a 3D model into 4D under strong coupling conditions.

To this end, we consider the freely acting type II asymmetric orbifold model. The theory is compactified on
\begin{align}
    \frac{T^6\times S^1}{\mathbb{Z}_N},
\label{eq:freely_acting}\end{align}
where the momentum lattice of $T^6$ is chosen to be a Lie algebra lattice.
The $\mathbb{Z}_N$ acts on $T^6$ as a twist, while acting on $S^1$ as a shift. The total $\mathbb{Z}_N$ action is freely acting, which lifts all the states in the twisted sectors.
Numerous models are constructed in this way.

An idea is to construct a 3d model whose spectrum is the same as the $S^1$ reduction of 4d one vector model. Subsequently, we hypothesize that the desired 4d model is realized at the strong coupling limit of the 3d model.

It is easy to realize such a 3d model. For instance, we take $\Gamma^{6,6}(E_6)+\Gamma^{1,1}$ lattice.
We choose the left and right twists on $\Gamma^{6,6}(E_6)$ as\footnote{See ~\cite{Carter1970,Bouwknegt:1988hn} for the conjugacy classes in the Weyl group of $E_6$.}
\begin{align}
    &\phi_L=\left(0,\frac{2}{3},\frac{2}{3}\right),
    &&\phi_R=\left(\frac{1}{9},\frac{2}{9},\frac{4}{9}\right).
\end{align}
This is the order $18$ twist by taking into account the fermions. 
It turns out that the untwisted sector spectrum is the desired one.
This can be seen by checking the spectrum before compactifying on $S^1$ (before adding $\Gamma^{1,1}$).
The 4d untwisted sector massless states are
\begin{align}
    \left[\mathbf{2}_v+2\mathbf{1}_\text{Scalar}+\mathbf{2}_\text{Spinor}\right]
    \times
    \left[\mathbf{2}_v\right]
    =(\text{4d $\mathcal{N}=2$ Gravity})+(\text{Vector}),
\label{Eq:4d_1V}\end{align}
where left hand side shows the left and right-moving states.
The massless spectrum of our 3d model is obtained by compactifying the above on $S^1$.
Moreover, all the twisted sectors are massive thanks to the shift on $S^1$ (see \eqref{eq:freely_acting}).

Let us study the behavior in the infinite distance limit of the model, which is either decompactification or tensionless string  limit according to the emergent string conjecture~\cite{Lee:2019wij,Lee:2019xtm}.
There are two non-compact scalars in our 3d model. One is the radius of freely acting $S^1$. Another is the scalar corresponding to the string coupling.
\begin{itemize}
    \item We can analyze the large radius limit using perturbative string theory. It turns out that when the radius of freely acting $S^1$ goes to infinity, the theory decompactifies to 4d theory which has more than eight supercharges. So, this limit is not interesting.  Another infinite distance is the small radius limit, but in this case it is easy to see that we get an enhanced supersymmetric model.
    \item 
    The strong coupling limit of this theory is not clear due to a lack of understanding regarding instanton corrections.
    It may be possible that the theory could decompactify into a 4D $\mathcal{N}=2$ model featuring a single vector multiplet in this limit. While we are unable to conclusively demonstrate this scenario, it is a viable option. 
\end{itemize}

Finally, in Table~\ref{tab:3d_models}, we provide a list of 3d models which have a chance to decompactify into the 4d model from rank $1$ to $5$. The first column is the label of the 3d models, and the second column is the rank of 4d model assuming that the strong coupling limit is the decompactification limit. When the 4d vector multiplet is compactificed on $S^1$, it becomes 3d vector multiplet. The number of (NSNS,RR) vector multiplet is shown in the third column. The fourth and last columns are the choice of the lattice and twist.
\renewcommand{\baselinestretch}{1.5}
\begin{longtable}{|c|c|c|c|c|c|}
\hline
3d models&
\begin{tabular}{c}
     \begin{tabular}{c}
          Rank of  \\
          4d model? 
     \end{tabular} 
\end{tabular} 
& 
\begin{tabular}{c}
     (NSNS, RR)  \\
     vector 
\end{tabular}  
& lattice ($+\Gamma^{1,1}$)
&   twist 
\\ \hline\hline

No.$1$ & $1$ &  $(1,0)$ & $\Gamma^{6,6}(E_6)$ & 
\begin{tabular}{c}
\\[-4mm]
$\phi_L=\left(0,\frac{2}{3},\frac{2}{3}\right)$, \\[+0.9mm]
$\phi_R=\left(\frac{1}{9},\frac{2}{9},\frac{4}{9}\right)$ \\[+1mm]
\end{tabular} 
\\[+0.3mm]\hline

No.$2$ & $3$ & $(1,2)$  & $\Gamma^{6,6}(A_6)$ & 
\begin{tabular}{c}
\\[-4mm]
$\phi_L=\left(0,\frac{1}{2},\frac{1}{2}\right)$,  \\[+0.9mm]
$\phi_R=\left(\frac{1}{7},\frac{2}{7},\frac{4}{7}\right)$
\\[+1mm]
\end{tabular} 
\\[+0.3mm]\hline

No.$3$ & \multirow{6}{*}{$5$} &  $(3,2)$ & $\Gamma^{6,6}(D_6)$ 
& 
\begin{tabular}{c}
     \\[-4mm]
     $\phi_L=\left(0,\frac{2}{3},\frac{2}{3}\right)$, \\[+0.9mm]
     $\phi_R=\left(0,\frac{1}{6},\frac{1}{2}\right)$  \\[+0.7mm]
\end{tabular}
\\ [+0.3mm]\cline{1-1}\cline{3-5}

No.$4$ & & $(1,4)$ & $\Gamma^{6,6}(D_6)$ &  
\begin{tabular}{c}
     \\[-4mm]
     $\phi_L=\left(0,\frac{3}{4},\frac{3}{4}\right)$, \\[+0.9mm]
     $\phi_R=\left(\frac{3}{2},\frac{1}{2},\frac{1}{2}\right)$  \\[+0.7mm]
\end{tabular}
\\[+0.3mm]\cline{1-1}\cline{3-5}

No.$5$ & & $(5,0)$ & $\Gamma^{6,6}(D_6)$ &
\begin{tabular}{c}
     \\[-4mm]
     $\phi_L=\left(0,\frac{1}{4},\frac{1}{4}\right)$,\\[+0.9mm]
     $\phi_R=\left(1,0,\frac{2}{3}\right)$  \\[+0.7mm]
\end{tabular}
\\[+0.7mm]\hline
\caption{3d models could be decompactified to 4d at the strong coupling. The shift is added to $\Gamma^{1,1}$ to make an orbifold freely-acting. The choice of the orbifold is not unique for each rank. }
\label{tab:3d_models}
\end{longtable}
\renewcommand{\baselinestretch}{1}

\section{Conclusion and discussion}\label{sec:conlusion}
In this paper, we constructed models without the geometric ``universal hypermultiplet" in 6, 5 and 4 dimensions.  Even though some examples of this is already in the literature we found more convincing evidence not only for their existence but more importantly for their transitions to geometric models with universal hypermultiplet.  Along the way we found that some geometric conditions, such as the Kodaira condition, can be violated by the non-geometric models we constructed.

We did not manage to construct no hyper and no vector theories with 8 supercharges in 5d or 4d theories.  Nor could we rule it out.  Moreover, we found that the only no hyper theories we could construct in 5d have even rank, starting with 2.  Furthermore, all no hyper theories in 4d have minimal rank 3. But we found theories in 3d which if they do have a 4d lift, then they would describe theories with smaller rank. These features we could not explain, and it remains to be seen whether they are Swampland conditions, or some consistent QG theories on the landscape can violate them.

\section*{Acknowledgements}

We thank Miguel Montero, Hee-Cheol Kim, Wati Taylor, Albrecht Klemm, Juven Wang, Gabi Zafir for useful discussions. We also thank Alon E. Faraggi, Stefan Groot Nibbelink, and Benjamin Percival for correspondence.
We thank the 2023 Summer Program of the Simons Center for Geometry and Physics for kind hospitality where part of this work was completed.
The work of YH is supported in part by MEXT Leading Initiative for Excellent Young Researchers Grant Number JPMXS0320210099. YH would like to thank Harvard University and University of Wisconsin-Madison for their hospitality during part of this work. The work of ZKB, HCT and CV is supported by a grant from the Simons Foundation (602883,CV), the DellaPietra Foundation, and by the NSF grant PHY-2013858. 
\appendix{}

\section{Orbifolds Review}\label{sec:review}
\subsection{Symmetric Orbifolds}
Consider string theory compactified on $T^d$. One can obtain a new theory by quotienting the $T^d$ by a isometry group $G$ of $T^d$ and get string theory compactified on the orbifold $T^d/G$ \cite{Dixon:1985jw,Dixon:1986jc}. This is called a \textit{symmetric orbifold} since both the left and right moving degrees of freedom of the string live in the same space.

An isometry $g\in G$ of $T^d$ lifts to that of $\mathbb R^d$, therefore it can be written as an orthogonal transformation $R$ together with a shift $v$
\begin{align}
    g = (R , v) \in \mathrm{O}(d)\ltimes \mathbb R^d.
\end{align}
Here, $R\in \mathrm{O}(d)$ can be assumed to be diagonal by complexifying the appropriate torus coordinates
\begin{align}
    Z^i &= \frac 1 {\sqrt{2}} (X^{2i} + i X^{2i+1}),\\
    Z^{i*} &= \frac 1 {\sqrt{2}} (X^{2i} - i X^{2i+1}),
\end{align}
where $i=1,\dots,\lfloor d/2 \rfloor$. If $d$ is odd, we keep $X^d$ as a real coordinate. In this basis, $R$ is a diagonal matrix
\begin{align}
    R = \mathrm{diag}(e^{2\pi i \phi_1},e^{-2\pi i \phi_1},\dots, e^{2\pi i \phi_{d/2 }},e^{-2\pi i \phi_{d/2}})
\end{align}
if $d$ is even and
\begin{align}
    R = \mathrm{diag}(e^{2\pi i \phi_1},e^{-2\pi i \phi_1},\dots, e^{2\pi i \phi_{\lfloor d/2\rfloor }},e^{-2\pi i \phi_{\lfloor d/2\rfloor}},\pm 1)
\end{align}
if $d$ is odd so that the eigenvalue on $X^d$ is $\pm 1$. The collection of the rotation angles
\begin{align}
    \phi = (\phi_1, \dots, \phi_{\lfloor d/2\rfloor })
\end{align}
is called the \textit{twist vector}. 

To define the action of rotations on superstrings, first we need $R\in \mathrm{SO}(d)$, and second we should lift $R$ to $\tilde R\in \mathrm{Spin}(d)$. For $R^N=I$ with $N$ odd, the choice of the lift corresponds to whether $g^N=(-1)^F$ or $g^N=e$, where $e\in G$ is the identity. In terms of the twist vector, the lift to $\mathrm{Spin}(d)$ corresponds to lifting the entries $\phi_i\in \mathbb R/\mathbb Z$ to $\tilde \phi_i \in \mathbb R /2\mathbb Z$. In particular, the lift choice corresponding to $\sum N \tilde \phi_i$ odd gives $g^N=(-1)^F$ and otherwise for $\sum N \tilde\phi_i$ even. For simplicity and by an abuse of notation, we write $\phi_i$ instead of $\tilde\phi_i$. Similarly, by $G$ we also mean its lift $\tilde G$ for supersymmetric strings.

To find the spectrum of the orbifold compactification, we start with the Hilbert space of the parent $T^d$ theory, the \textit{untwisted sector} $\mathcal H_e$. We first only consider the states $\ket{\psi}\in \mathcal H_e$ that are invariant under the action of $g\in G$
\begin{align}
    g\cdot \ket{\psi} = \ket{\psi}.
\end{align}
Then the partition function of the untwisted sector can be computed by an insertion of the projector $\pi_G=\frac{1}{|G|}\sum_{g\in G}g$ as
\begin{align}
    Z_{e} = \frac{1}{|G|}\sum_{g\in G} \mathrm{Tr}_{\mathcal H_e}(gq^{L_0-c/24}\bar q^{\bar L_0-\bar c/24}) = \frac{1}{|G|}\sum_{g\in G} Z_e^g.
\end{align}
Here, $Z_e^g$ are \textit{partial traces} corresponding to traces in the untwisted sector $\mathcal H_e$ with an insertion of $g$.

It turns out that the untwisted sector alone is not modular invariant under $\tau\mapsto -1/\tau$. This suggests that the spectrum for the orbifold compactification includes more than just the untwisted sector. Indeed, one can have strings with boundary conditions that close up to a group action by $g\in G$ as
\begin{align}
    X^i(\sigma+2\pi) = g\cdot X^i(\sigma).
\end{align}
These states correspond to the \textit{$g$-twisted sectors}, with the untwisted sector corresponding to the identity $g=e$. In the $g$-twisted sector, we only take the states that are invariant under the centralizer $C_G(g):=\{x\in G\mid gx=xg\}$. Denoting by $\mathcal H_g$ the $g$-twisted sector, the partition function of the $g$-twisted sector is
\begin{align}
    Z_g = \frac{1}{|G|}\sum_{h\in C_G(g)} \mathrm{Tr}_{\mathcal H_g}(hq^{L_0-c/24}\bar q^{\bar L_0 - \bar c/24}) = \frac{1}{|G|}\sum_{h\in C_G(g)} Z_g^h.
\end{align}
Similar to before, $Z_g^h$ denotes the partial trace in the $g$-twisted sector with an $h$ insertion.

The total partition function is
\begin{align}
    Z =\sum_{g\in G} Z_g = \frac{1}{|G|}\sum_{g\in G}\sum_{h\in C_G(g)} Z_g^h,
\end{align}
which finally is modular invariant.

For our purposes, it is enough to consider \textit{cyclic orbifolds}
\begin{align}
    G = \langle g \rangle \cong \mathbb Z_N,
\end{align}
where the isomorphism is given by $g^n\mapsto n$. The centralizer is simply the whole group $C_G(g^n) = G$ and we have
\begin{align}
    Z = \frac 1 N \sum_{g,h=0}^N Z_g^h.
\end{align}
\subsection{Asymmetric Orbifolds}\label{sec:asymmetric}
Since left and right moving degrees of freedom of strings are decoupled, we can consider orbifolding by a symmetry $g$ that acts differently on the left and the right. Such an orbifold has no target space interpretation and is called an \textit{asymmetric orbifold} \cite{Narain:1986qm,Narain:1990mw}.

In particular, string theory compactified on $T^d$ is characterized by an even unimodular lattice called the \textit{Narain lattice} $\Gamma^{d+x,d}$ with signature $(d+x,d)$, where $x=16$ for heterotic strings and $x=0$ otherwise \cite{Narain:1985jj, Narain:1986am}. We only consider the automorphisms of the Narain lattice that decompose to left and right components as
\begin{align}
    (R_L;R_R)\in \mathrm{O}(d+x)\times \mathrm{O}(d) \subset \mathrm{Aut}(\Gamma^{d+x,d}).
\end{align}
Coupling the automorphism with shifts on the left and right $v_L,v_R \in \mathbb Q \otimes \Gamma^{d+x,d}$, we obtain an action
\begin{align}
    g = (R_L,v_L;R_R,v_R).
\end{align}

For $g^N=1$ with even $N$, a condition we impose is
\begin{align}
    pg^{N/2}p=0
    \quad\quad \text{mod $2$} 
\label{Eq:consistency_even}\end{align}
for all $p\in\Gamma^{d+x,d}$.\footnote{The cases in which condition \eqref{Eq:consistency_even} is not satisfied involve subtleties, see \cite{Harvey:2017rko}.}

We also define the invariant sublattice under $g^m$ as
\begin{align}
    I_m:=\mathrm{Fix}_{g^m} (\Gamma^{d+x,d}) = \{p\in \Gamma^{d+x,d}\mid g \cdot p=p\}.
\end{align}
For notational simplicity, we let $I:=I_1$.

By complexifying the left and right movers similarly to the symmetric case, we obtain two twist vectors, $\phi_L$ and $\phi_R$, associated to $R_L$ and $R_R$. The discussion concerning the lifts to the $\mathrm{Spin}$ group follow similarly for the left and right movers if they are supersymmetric. Each component of $\phi_{L,R}$ satisfies $-1\leq\phi_i<1$.

In summary, the data required to construct an asymmetric orbifold consists of the Narain lattice $\Gamma^{d+x,d}$, the twist vectors $\phi_L,\phi_R$, and the shift vector $v=(v_L;v_R)$.

\subsubsection{Type II}
We consider the case where the momentum lattice $\Gamma^{d,d}$ is a Lie algebra lattice. Namely,
\begin{align}
    \Gamma^{d,d}(\mathfrak g):=\{(p_L;p_R)|p_L\in\Lambda_W( \mathfrak g),p_R\in\Lambda_W(\mathfrak g),p_L-p_R\in\Lambda_R(\mathfrak g)\}.
\end{align}
This lattice is even as $p_L$ and $p_R$ belong to the same conjugacy class.
Moreover, the lattice is self-dual since 
\begin{align}
    &p_L\cdot q_L-p_R\cdot q_R=0\quad\text{mod }1,
    &&(p_L,p_R),(q_L,q_R)\in\Gamma^{d,d}.
\end{align}

To compute the spectrum for type II strings on asymmetric $\mathbb Z_N$ orbifolds, we consider the weight vectors $r_L$ and $r_R$ of $\mathrm{SO}(8)$ and the mass formulae in the $m$-th twisted sector:
\begin{align}
    &H_{L,m}=L_{0,m}-\frac{1}{2},\label{Eq:H_LR}
    \quad\quad\quad\quad H_{R,m}=\overline{L}_{0,m}-\frac{1}{2},\\
    &L_{0,m}=N_B+\frac{(r_L+m\phi_L)^2}{2}+\frac{(p_L+mv_L)^2}{2}+E_0,\\
    &\tilde{L}_{0,m}=\tilde{N}_B+\frac{(r_R+m\phi_R)^2}{2}+\frac{(p_R+mv_R)^2}{2}+\tilde{E}_0\\
    &E_0=\sum_i\frac{1}{2}|[m \phi_{L,i}]|(1-|[m \phi_{L,i}]|),
    \quad\quad\quad\quad \tilde{E}_0=\sum_i\frac{1}{2}|[m \phi_{R,i}]|(1-|[m \phi_{R,i}]|).
\end{align}
Here, $[m\phi_i]$ is defined as $-1\leq[m\phi_{i}]<1$ with $m\phi_i=[m\phi_i]+2l_i$ and $l_i\in\mathbb{Z}$. The momenta $(p_L,p_R)$ in the $m$-th twisted sector live in the dual lattice $I_m^*$. The vector $r_{L,R}$ is
\begin{align}
    r_L&=
    \begin{cases}
    (n_1,n_2,n_3,n_4), &\sum n_i=\text{odd},\quad\text{NS-sector}\quad \mathbf{8}_v,\\ \\
    \left(n_1+\dfrac{1}{2},n_2+\dfrac{1}{2},n_3+\dfrac{1}{2},n_4+\dfrac{1}{2}\right), &\sum n_i=\text{odd},\quad\text{R-sector}\quad \mathbf{8}_c,
    \end{cases}\label{Eq:r_L}\\
    r_R&=
    \begin{cases}
    (n_1,n_2,n_3,n_4), &\sum n_i=\text{odd},\quad\text{NS-sector}\quad \mathbf{8}_v,\\ \\
    \left(n_1+\dfrac{1}{2},n_2+\dfrac{1}{2},n_3+\dfrac{1}{2},n_4+\dfrac{1}{2}\right), &\sum n_i=\text{odd},\quad\text{R-sector (IIB)}\quad \mathbf{8}_c,\\ \\
    \left(n_1+\dfrac{1}{2},n_2+\dfrac{1}{2},n_3+\dfrac{1}{2},n_4+\dfrac{1}{2}\right),&\sum n_i=\text{even},\quad\text{R-sector (IIA)}\quad \mathbf{8}_s.
    \end{cases}
\label{Eq:r_R}\end{align}

The necessary and sufficient condition for modular invariance is the level matching condition \cite{Vafa:1986wx}
\begin{align}
    N(H_{L,m}-H_{R,m})\in\mathbb{Z}
\label{Eq:level_matching}\end{align}
for any $N_B, \tilde{N}_B, r_{L,R}$ and $p_{L,R}$.
For example, the necessary conditions to satisfy the level matching are
\begin{align}
    &\frac{(v^*)^2}{2}\in\frac{\mathbb{Z}}{N},
    &&\sum_i \phi_{L,i}\in\frac{2\mathbb{Z}}{N},
    &&\sum_i \phi_{R,i}\in\frac{2\mathbb{Z}}{N}.
\label{Eq:consistency}\end{align}
Here, $v^*$ is the orthogonal projection of $v$ to the space spanned by $I$.

\subsubsection{Heterotic}
For heterotic strings, the Narain lattice is $\Gamma^{d+16,d}$. Since the Narain lattice is asymmetric, we have to use different Lie algebra lattices on the left and the right. For the left and right Lie algebra lattices to be compatible, we need their conjugacy class groups to be isometric. 

In particular, define the \textit{glue group}
\begin{align}
    \mathcal D(\mathfrak g) := \Lambda_W(\mathfrak g)/\Lambda_R(\mathfrak g)
\end{align}
with $p\in \Lambda_W(\mathfrak g)$ projecting to $[p]\in \mathcal D(\mathfrak g)$.
Endow the glue group with the quadratic from $q$ of $\Lambda_W(\mathfrak g)$ as
\begin{align}
    \bar q([v]) := q(v) \mod{2},\qquad v\in \Lambda_W(\mathfrak g).
\end{align}
Then, given $\mathfrak h$ and an isometry,
\begin{align}
    \psi: \mathcal D(\mathfrak g) \to \mathcal D(\mathfrak h),
\end{align}
the lattice
\begin{align}
    \Gamma^{d+16,d}(\mathfrak g, \mathfrak h):=\{(p_L;p_R)|p_L\in\Lambda_W( \mathfrak g),p_R\in\Lambda_W(\mathfrak h),\psi([p_L])=[p_R]\}
\end{align}
is even and unimodular.

Note that since $\mathcal D(E_8)$ is trivial, we get a simplification for $E_8\times E_8$ constructions as
\begin{align}
    \Gamma^{d+16,d}(E_8\times E_8\times \mathfrak g, \mathfrak g) = 2\Gamma^{8,0}(E_8) \oplus \Gamma^{d,d}(\mathfrak g).
\end{align}

Computing the spectrum for the heterotic string is identical to type II for the right movers, whereas the left movers are bosonic and involve different formulae. In the $m$-th twisted sector, we have
\begin{align}
    &H_{L,m}=N_B+\frac{(P+mV)^2}{2}+E_0-1,\\
    &H_{R,m}=\tilde{N}_B+\frac{(r_R+m\phi_R)^2}{2}+\frac{(p_R+mv_R)^2}{2}+\tilde{E}_0-\frac{1}{2}.
\end{align}

The bosonic zero point energy $E_0$ for $m$-th twisted sector is
\begin{align}
    E_0=
    \begin{dcases}
        \sum_{i=1}^{8+\lfloor d/2\rfloor}\frac{1}{2}|[m \phi_{L,i}]|(1-|[m \phi_{L,i}]|)\quad
       & \text{for $\det R_L=1$},\\
        \frac{1}{16}+\sum_{i=1}^{7+\lfloor d/2\rfloor}\frac{1}{2}|[m \phi_{L,i}]|(1-|[m \phi_{L,i}]|)
        \quad &\text{for $\det R_L=-1$}.
    \end{dcases}
\end{align}
The level-matching conditions are similar to those of type II.

\subsection{Freely acting orbifolds}\label{sec:shift}
The orbifold obtained by a group $G$ without fixed points is termed a \textit{freely acting orbifold}. The most useful feature of these orbifolds is that their twisted sectors can be massed up.

The technique we extensively use involves orbifolding on $T^d$ coupled with a shift on an additional $S^1$, compactifying overall to $9-d$ dimensions. The advantage of this construction is that except at special $S^1$ radii, all twisted sectors become massive. Essentially, freely acting orbifolds enable us to project out a significant portion of the massless spectrum without introducing massless states in the twisted sectors.

Consider an orbifolding action $g$ of order $N$ on $T^d$. Further compactify on an additional circle $S^1$ with radius $r$, and couple this action $g$ with a shift $2\pi r/N$ on the circle.

Since the $S^1$ remains invariant under the overall orbifolding action, we have
\begin{align}
    I \supset \Gamma^{1,1} = \left\{\frac 1 2(n/r + w r,n/r - w r)|n,w\in\mathbb Z\right\}\,.
\end{align}
A $2\pi R/N$ shift on $S^1$ corresponds to a pure winding mode vector $(m=0,w=1)$ in $\Gamma^{1,1}$ divided by $N$. Therefore the shift vector is
\begin{align}
    v = \frac 1 {2N} (r,-r)\,.
\end{align}

We will now prove that all twisted sectors are massive for large enough $r$. We will show this for type II strings, but the proof for heterotic strings is similar. Intuitively, as $r$ increases, twisted sector strings become longer and thus gain mass.

To prove that the $m$-th twisted sector with $m\neq 0$ is massive, we will show that $r$ can be chosen large enough such that
\begin{align}
    H_{L,m} \supset \frac{(p_L+mv_L)^2}{2} + E_0 - \frac 1 2 >0 
\end{align}
or
\begin{align}
    H_{R,m} \supset \frac{(p_R+mv_R)^2}{2} + \tilde E_0 - \frac 1 2 >0
\end{align}
holds for all $(p_L;p_R)$. Essentially, we will prove that either the left or right moving ground state is always massive.

Assume without loss of generality that there exists $p_L=\frac 1 2(\frac n r + w r)$ such that
\begin{align}\label{eq:pL-inequality}
    0\leq \frac{(p_{L}+m v_{L})^2}{2} = \frac{(\frac m {2N} r + \frac n r + w r)^2}{2} \leq  \frac 1 2- E_0 
\end{align}
for some $n,w\in \mathbb Z$. This implies potential massless states in the left-moving sector. We claim that $H_{R,m}>0$. If $\tilde E_0 - 1/2>0$, we are done, so we assume $\tilde E_0-1/2\leq 0$. 

Taking the positive square root of \eqref{eq:pL-inequality},
\begin{align}\label{eq:positivesqrt}
    \frac{|\frac m {2N} r + \frac n r + w r|}{\sqrt{2}} \leq  \sqrt{\frac 1 2- E_0}.
\end{align}
For the right movers, we have
\begin{align}
    \frac{(p_{R}+m v_{R})^2}{2} &= \frac{(-\frac m {2N} r + \frac n r - w r)^2}{2}.
\end{align}
Taking square root and using the reverse triangle inequality,
\begin{align}
    \frac{|-\frac m {2N} r + \frac n r - w r|}{\sqrt{2}} 
    &\geq \left|\frac{2|\frac m {2N} r  + w r|}{\sqrt{2}} -\frac{|\frac m {2N} r + \frac n r + w r|}{\sqrt{2}} \right|.
\end{align}
Using \eqref{eq:positivesqrt} we get
\begin{align}\label{eq:lastineq}
    \frac{2|\frac m {2N} r  + w r|}{\sqrt{2}} -\frac{|\frac m {2N} r + \frac n r + w r|}{\sqrt{2}} \geq \frac{2|\frac m {2N} r  + w r|}{\sqrt{2}} -\sqrt{\frac 1 2 - E_0}.
\end{align}
We can choose $r$ large enough so that the RHS of \eqref{eq:lastineq} is strictly larger than $\sqrt{\frac 1 2 - \tilde E_0}$. We conclude that
\begin{align}
    \frac{(p_R+m v_R)^2}{2} > \frac 1 2 - \tilde E_0,
\end{align}
so $H_{R,m}>0$ as claimed.

\section{Landscape of 6d Supergravity}\label{app:6dreview}
\subsection{Review of 6d (1,0) supergravity}
The 6d $(1,0)$ supergravity has a massless spectrum given by representation of the little group $SO(4)\cong SU(2)\times SU(2)$.

\begin{table}[h!]
\begin{tabular}{|c|c|}
\hline & \\
       & \textbf{Fields in  $SO(4)\cong SU(2)\times SU(2)$}  \\ & \\ \hline
\textbf{Gravity Multiplet (G)} & \begin{tabular}[c]{@{}c@{}}$\underbrace{(\mathbf{3},\mathbf{3})}_{g_{\mu \nu}}  +2\underbrace{(\mathbf{2},\mathbf{3})}_{2{\psi_\mu^+}} +\underbrace{(\mathbf{1},\mathbf{3})}_{B_{\mu \nu}^+}$ \\     \end{tabular} \\ \hline
\textbf{Tensor Multiplet (T)}  & \begin{tabular}[c]{@{}c@{}}$\underbrace{(\mathbf{3},\mathbf{1})}_{B^-_{\mu \nu}}    +2\underbrace{(\mathbf{2},\mathbf{1})}_{2 \psi^-}+\underbrace{(\mathbf{1},\mathbf{1})}_{\phi}$\end{tabular}                   \\ \hline
\textbf{Vector Multipet (V) }   & \begin{tabular}[c]{@{}c@{}}$\underbrace{(\mathbf{2},\mathbf{2})}_{A_\mu}    +2\underbrace{(\mathbf{1},\mathbf{2})}_{2\lambda^+}$\end{tabular}                                                                            \\ \hline
\textbf{Hypermultiplet (H)}    & \begin{tabular}[c]{@{}c@{}}$4\underbrace{(\mathbf{1},\mathbf{1})}_{4 \phi}    +2\underbrace{(\mathbf{2},\mathbf{1})}_{\chi^-}$\end{tabular}                    \\ \hline
\end{tabular}
\end{table}

In terms of the bosonized momentum $r$ (see Appendix~\ref{sec:asymmetric}), the representations of $SU(2)\times SU(2)$ are
\begin{align}
    &(\mathbf{1},\mathbf{1}): r=(\#,\#,0,0),
    \nonumber\\
    &(\mathbf{2},\mathbf{1}): r=\left(\#,\#,\underline{\frac{1}{2},-\frac{1}{2}}\right),
    \nonumber\\
    &(\mathbf{1},\mathbf{2}): r=\left(\#,\#,\pm\frac{1}{2},\pm\frac{1}{2}\right),
    \nonumber\\
    &(\mathbf{2},\mathbf{2}): r=(\#,\#,\underline{\pm1,0}),
\end{align}
where the underline indicates the permutation.
See \cite{Erler:1993zy,Aldazabal:1997wi} for earlier work on 6d heterotic string.

The tensor multiplet moduli space locally takes the form  $SO(1, T)/SO(T)$ , and can
be parameterized by a vector $j$
 in the space $\mathbb{R}^{1,T}$ of positive norm $j\cdot j >0$,  representing the positivity of the metric on the
moduli space. 

 The chiral fields of those multiplets contribute to the anomalies produced in such a  theory characterized by an 8-form anomaly polynomial $I_8$. Such anomalies can be cancelled by the Green-Schwarz-Sagnotti mechanism \cite{Sagnotti:1992qw} if the anomaly polynomial $I_8$ factorizes as 
 \begin{eqnarray}
 I_8(R,F)=\frac{1}{2}\Omega_{\alpha\beta}X^\alpha_4 X^\beta_4, \ \ X_4^\alpha=\frac{1}{2}a^\alpha trR^2 +\sum_ib_i^\alpha \frac{2}{\lambda_i}trF_i^2
 \end{eqnarray}
 where $a^\alpha, b_i^\alpha$ are vectors in $\R^{1,T}$, $\Omega_{\alpha \beta }$ is the metric on this space and  $ \lambda_i $ are normalization factors of the gauge groups $G_i$.
 The anomaly factorization conditions for gravitational, gauge and mixed anomalies are summarized as follows:

 \begin{itemize}
 	\item ${\label{eqn:R4} R^4:  \ \ H-V=273-29T}$
 	\item$ {\label{eqn:F4}F^4: \ \ 0=B^i_{Adj}-\sum n_R^i B^i_R}$
 	\item ${\label{eqn:R22}(R^2)^2: \  a\cdot a=a^\alpha\Omega_{\alpha \beta }a^\beta  =9-T}$
 	\item ${\label{eqn:F2R2} F^2R^2: \ a\cdot b_i=a^\alpha\Omega_{\alpha \beta }b_i^\beta   =\frac{1}{6}\lambda_i (A^i_{Adj}-\sum_Rn_R^iA^i_R)} $
 	\item ${\label{eqn:F22}(F^2)^2: \  b_i \cdot b_i =b_i^\alpha\Omega_{\alpha \beta }b_i^\beta  =\frac{1}{3}\lambda_i^2 (\sum_R n_R^iC^i_R-C^i_{Adj}) }$
 	\item ${\label{eqn:F2F2}F^2_iF^2_j: \ b_i \cdot b_j=b_i^\alpha\Omega_{\alpha \beta }b_j^\beta  = \sum_{R,S}\lambda_i \lambda_jn_{RS}^{ij}A^i_RA^j_S  \ \ \ i\neq j }$
 	 \end{itemize}
 	where $H, V, T$  denote the number of hypermultiplets, vectors multiplets and tensor multiplets in the theory respectively. The number  $n_R^i$ represents the number of hypermultiplets in the representation $\textbf{R}$ of the gauge group $G_i$ and $A_R^i,B_R^i,C_R^i$ are the following group theory coefficients:
 	\begin{eqnarray}
 	tr_{\text{R}}F^2=A_R trF^2, \quad	tr_{\text{R}}F^4=B_R trF^4+C_R(trF^2)^2
 	\end{eqnarray} 
the values of some of those coefficients for various representations and the normalization factors $\lambda_i$ are summarized in Table \ref{Tab:Casimirs}.
In addition, as shown in \cite{Kumar_2010} the vectors  $a^\alpha,b_i^\alpha \in \R^{1,T}$ are constrained to have integer inner products $ a\cdot a , a\cdot b_i , b_i \cdot b_j \in \Z$ with respect to the bilinear form $\Omega_{\alpha \beta}$, we call this the anomaly lattice. The gauge coupling of the group $G_i$ is given by $g_i^{-2}=j\cdot b_i$. Positivity of the kinetic terms and the Gauss-Bonnet term implies $j\cdot b_i\geq 0 $ and $j\cdot a\leq 0$ \cite{Kumar:2010ru}.
\begin{table}[h!]
\begin{tabular}{|c|c|c|c|c|c|}
\hline
        & $\lambda$ & $A_{adj}$ & $C_{adj}$ & $A_{R}$ & $C_{R}$         \\ \hline
$SU(2)$ & 1         & 4         & 1/2       & $\mathbf{2}=1$        & $\mathbf{2}=8$                \\ \hline
$SU(3)$ & 1         & 6         & 9         & $\mathbf{3}=1$        & $\mathbf{3}=1/2$              \\ \hline
$SO(8)$ & 2         & 6         & 3         & $\mathbf{8}_{s,c,v}=1$ & $\mathbf{8}_s,\mathbf{8}_c=3/2^3,\mathbf{8}_v=0$ \\ \hline
$E_6$   & 6         & 4       & 1/2       &  $\mathbf{27}=1$    &  $\mathbf{27}=1/12$            \\ \hline
$E_7$   & 12        & 3         & 1/6       & 1       & 1/24            \\ \hline
$E_8$   & 60        & 1         & 1/100     &         &                 \\ \hline
\end{tabular}
\caption{Various group theoretic invariants.}
\label{Tab:Casimirs}
\end{table}

The charges $a^\alpha, b^\alpha_i$ span the anomaly lattice $\Lambda$ which is contained in
the full string lattice of the 6d theory. The anomaly
lattice is required to have a unimodular embedding into a self-dual lattice \cite{Seiberg:2011dr}. Furthermore, the existence of the
two-form fields 
implies the existence of string sources in accordance with the
hypothesis that the spectrum of a gravitational theory needs to be complete \cite{Banks:2010zn,Polchinski:2003bq}. Such considerations provide various constraints on the possible 6d theories as demonstrated in \cite{Kim:2019vuc,Tarazi:2021duw}.

For example, a  non-instantonic BPS string with charge $Q$ satisfies 
\begin{equation}
    \sum_{G_i}c_{G_i}\leq c_L=3Q\cdot Q-9a\cdot Q+2
\end{equation}
where $G_i$ is some bulk gauge symmetry that appears as a current algebra on the string worldsheet with $c_{G_i}$ the corresponding central charges.

In the F-theory compactifications on a elliptic Calabi-Yau threefold the massless spectrum can be determined by the Hodge numbers: 

\begin{align}
    & h^{1,1}(CY_3)=r_G+T  & h^{2,1}(CY_3) =H_{\text{neutral}}-1
\end{align}

This implies that such theories have at least one neutral hypermultiplet, known as the "universal" hypermultiplet. In \autoref{sec:6d} we demonstrate the existence of theories with no neutral hypermultiplets corresponding to "$h^{2,1}=-1$" which are realized by   non-geometric constructions.

The map between the anomaly coefficients and  F-theory is:

\begin{align}
 &   a\to K_B \\
& b_i \to C_i \\
& j \to J_B
\end{align}
where $K_B$ is the canonical divisor of the base, $C_i$ effective and irreducible divisors that support that gauge algebra $G_i$ and $J_B$ the K\"ahler class of the base.
Therefore, the gauge coupling is identified with the volume of the corresponding divisor as $g_i^{-2}=Vol_{J_B}(C_i)$, this also controls  the tension of the string with charge $b_i$ which corresponds to the D3 brane wrapping $C_i$.

Additionally, a  consistent elliptic fibration  over a base $B$ needs to satisfy the\textit{ Kodaira condition} \cite{Kumar_2010} given by 
\begin{equation}
    -12K_B=\sum_i \nu_i C_i+Y
\end{equation}
where $\nu_i$ are the singularity multiplicities and $Y$ the residual divisors which is the sum of effective divisors associated to non-contractible curves.

\subsection{More details on constructions}\label{app:AOdetails}
\subsubsection*{type II Model 1}
We consider the type IIB on $T^4$ asymmetric orbifold.\footnote{The same spectrum is obtained for type IIA string.} The momentum lattice takes the form
\begin{align}
    \Gamma^{4,4}(D_4)=\{(p_L,p_R)|p_L\in\Lambda_W(D_4),p_R\in\Lambda_W(D_4),p_L-p_R\in\Lambda_R(D_4)\}.
\end{align}
at a special point in the Narain lattice. Here, $\Lambda_R(D_4)$ and $\Lambda_W(D_4)$ are the root and weight lattices of $D_4$, respectively. 
In terms of the orthonormal basis, an element of $\Lambda_R(D_4)$ lattice is written as $(n_1,n_2,n_3,n_4)$ with $\sum n_i\in2\mathbb{Z}$ and $n_i\in\mathbb{Z}$, while an element of $\Lambda_W(D_4)$ lattice is written as $(n_1,n_2,n_3,n_4)$ or $(n_1+1/2,n_2+1/2,n_3+1/2,n_4+1/2)$ with $n_i\in\mathbb{Z}$.
Next, we perform an asymmetric $\mathbb{Z}_2$ twist:
\begin{align}
    &\phi_L=\left(\frac{1}{2},\frac{1}{2}\right),
    &&\phi_R=(1,0),
\end{align}
where the eigenvalues of the twist are $(e^{2\pi i (\phi_L)_1},e^{-2\pi i (\phi_L)_1},e^{2\pi i (\phi_L)_2},e^{-2\pi i (\phi_L)_2})$ for the left-moving sector and $(e^{2\pi i (\phi_R)_1},e^{-2\pi i (\phi_R)_1},e^{2\pi i (\phi_R)_2},e^{-2\pi i (\phi_R)_2})$ for the right-moving sector. The twist $\phi_R$ corresponds to $(-1)^{F_R}$, where $F_R$ is the right-moving fermion number. Consequently, all the right-moving supersymmetry is broken. On the other hand, half of the left-moving supersymmetry is preserved as the twist $\phi_L$ belongs to $SU(2)$.
This satisfies the mod 2 condition in \cite{Narain:1986qm} as $pg p=-p_L^2-p_R^2$ is even for $p_L-p_R\in\Lambda_R(D_4)$. The invariant lattice $I$ and the dual lattice $I^*$  are
\begin{align}
    &I=\{\,(0,p_R)\,|\,p_R\in\Lambda_R(D_4)\,\},
    &&I^*=\{\,(0,p_R)\,|\,p_R\in\Lambda_W(D_4)\,\}.
\end{align}
The spectrum in the untwisted and twisted sectors is computed as follows.

\begin{itemize}
    \item Untwisted sector
    
    The untwisted massless spectrum is computed as in Table~\ref{Tab:Z2_no_neutral_untwisted}.
\begin{longtable}{|c|ccc|}\hline \hline
  Phase    &   Left & Right & Total            
  \\\hline\hline\\[-1em]
$0$   &  $(\mathbf{2},\mathbf{2})+2(\mathbf{1},\mathbf{2})$ & $(\mathbf{2},\mathbf{2})+4(\mathbf{1},\mathbf{1})$  &
\begin{tabular}{c}
$\underbrace{(\mathbf{3},\mathbf{3})+2(\mathbf{2},\mathbf{3})+(\mathbf{1},\mathbf{3})}_{G}$ \\
$+\underbrace{(\mathbf{3},\mathbf{1})+2(\mathbf{2},\mathbf{1})+(\mathbf{1},\mathbf{1})}_{T}$\\
$+\underbrace{4(\mathbf{2},\mathbf{2})+8(\mathbf{1},\mathbf{2})}_{4V}$
\end{tabular}
\\[+0.5em]\hline\\[-1.em]

$\dfrac{1}{2}$   & 
$2(\mathbf{2},\mathbf{1})+4(\mathbf{1},\mathbf{1})$ 
& 
$2(\mathbf{2},\mathbf{1})+2(\mathbf{1},\mathbf{2})$ 
& 
 \begin{tabular}{c} 
 $\underbrace{4(\mathbf{2},\mathbf{2})+8(\mathbf{1},\mathbf{2})}_{4V}$\\
 $+\underbrace{4(\mathbf{3},\mathbf{1})+8(\mathbf{2},\mathbf{1})+4(\mathbf{1},\mathbf{1})}_{4T}$
 \end{tabular}
\\[+0.6em]\hline

\caption{
The untwisted sector spectrum in Model 1.
}
\label{Tab:Z2_no_neutral_untwisted}
\end{longtable}
In Table~\ref{Tab:Z2_no_neutral_untwisted}, the representation under the 6d little group $SO(4)\simeq SU(2)\times SU(2)$ is listed. 
From Table~\ref{Tab:Z2_no_neutral_untwisted}, the untwisted spectrum is
\begin{align}
    G+5T+8V,
\end{align}
where $G, T$, and $V$ are the gravity, tensor, and vector multiplets, respectively.

    \item Twisted sector

    The degeneracy factor~\cite{Narain:1986qm} is
    \begin{align}
    \frac{\prod_{i=1}^2 (2\sin(\pi \phi_{Li}))}{V_I}=2,
    \end{align}
    where $V_I=2$ is the volume of the lattice $I$.
    From Eq.~\eqref{Eq:H_LR}, the left-mover massless states are
    \begin{align}
        &(\mathbf{2},\mathbf{1}):
        r_L=\left(-\frac{1}{2},-\frac{1}{2},\underline{-\frac{1}{2},\frac{1}{2}}\right),
        \\
        &2(\mathbf{1},\mathbf{1}):
        r_L=(\underline{-1,0},0,0).
    \end{align}
    On the other hand, the right-mover massless states are
    \begin{align}
        &2(\mathbf{2},\mathbf{1}):
        r_R=\left(-\frac{1}{2},-\frac{1}{2},\underline{-\frac{1}{2},\frac{1}{2}}\right),\quad
        \left(-\frac{3}{2},\frac{1}{2},\underline{-\frac{1}{2},\frac{1}{2}}\right),\quad
        p_R=0,\\
        &2(\mathbf{1},\mathbf{2}):
        r_R=\left(-\frac{1}{2},\frac{1}{2},\pm\frac{1}{2},\pm\frac{1}{2}\right),
        \quad\left(-\frac{3}{2},-\frac{1}{2},\pm\frac{1}{2},\pm\frac{1}{2}\right),\quad
        p_R=0,\\
        &24(\mathbf{1},\mathbf{1}):
        r_R=(-1,0,0,0), \nonumber\\
        &p_R=\left(\underline{\pm1,0,0,0}\right),
        \pm\left(\frac{1}{2},\frac{1}{2},\frac{1}{2},\frac{1}{2}\right),
        \pm\left(\underline{-\frac{1}{2},\frac{1}{2},\frac{1}{2},\frac{1}{2}}\right),
        \left(\underline{-\frac{1}{2},-\frac{1}{2},\frac{1}{2},\frac{1}{2}}\right).
    \label{Eq:charged_hyper}\end{align}
    Note that $p_R\in I^*=\Lambda_W(D_4)$. There are 24 states with $p_R^2=1/2$ corresponding to $\mathbf{8}_v$, $\mathbf{8}_s$, and $\mathbf{8}_c$ representations of $D_4$.

    Taking into account the degeneracy factor $2$, the twisted sector spectrum is
    \begin{align}
        &2\left[(\mathbf{2},\mathbf{1})+2(\mathbf{1},\mathbf{1})\right]
        \times
        \left[2(\mathbf{2},\mathbf{1})+2(\mathbf{1},\mathbf{2})+24(\mathbf{1},\mathbf{1})\right]
        \nonumber\\
        &=
        \underbrace{4(\mathbf{3},\mathbf{1})+8(\mathbf{2},\mathbf{1})+4(\mathbf{1},\mathbf{1})}_{4T}
        +\underbrace{4(\mathbf{2},\mathbf{2})+8(\mathbf{1},\mathbf{2})}_{4V}
        +\underbrace{48(\mathbf{2},\mathbf{1})+96(\mathbf{1},\mathbf{1})}_{24H_c}.
    \end{align}
    All the hypermultiplets arise from states with $p_R\neq0$, as can be observed from Eq.~\eqref{Eq:charged_hyper}. 
    Since the lattice $I^*$ can be regarded as a charge lattice of $U(1)^4$, all the hypermultiplets carry charges.

    \item Full spectrum

    By combining the untwisted and twisted sectors, the full massless spectrum is
    \begin{align}
    G+9T+12V+24H_c.
    \end{align}
    The gauge group is $U(1)^{12}$. The charges of the hyper are
    \begin{align}
        &\left(\underline{\pm1,0,0,0},0^8\right)
        +\left(\pm\frac{1}{2},\pm\frac{1}{2},\pm\frac{1}{2},\pm\frac{1}{2},0^8\right)
        \nonumber\\
        &+\left(\underline{\pm\frac{1}{2},\mp\frac{1}{2},\mp\frac{1}{2},\mp\frac{1}{2}},0^8\right)
        +\left(\underline{-\frac{1}{2},-\frac{1}{2},\frac{1}{2},\frac{1}{2}},0^8\right)
    \label{eq:model1_charge}\end{align}
    under $U(1)^{12}$. The gravitational anomaly-free condition $H-V=273-29T$ is satisfied. 
\end{itemize}

\subsubsection*{Heterotic model 2}
We consider the heterotic string theory compactified on $T^4$, where the momentum lattice is
\begin{align}
    \Gamma^{4,4}(D_4)+2\Gamma^{8,0}(E_8).
\end{align}
Then, we perform the following $\mathbb{Z}_2$ twist and shift,\footnote{Just from the level matching condition alone, $V_L=0$ is allowed. However, in this case, extra gravitinos appear from the twisted sector, and the theory becomes 6d $\mathcal{N}=(1,1)$.}
\begin{align}
    &\phi_L=(0,0),
    \quad\quad\quad\quad\quad\quad\phi_R=\left(\frac{1}{2},\frac{1}{2}\right),
    \nonumber\\
    &V_L=\frac{1}{2}(1^2,0^6;1^2,0^6).
\label{eq:model2}\end{align}
Here $T^4$ is chosen as $\Gamma^{4,4}(D_4)$ lattice, and $V_L$ is the shift in $E_8\times E_8$ lattice.\footnote{There are no shifts in $T^4$ direction.} As the twist $\phi_R$ sits in $SU(2)$, the half of the right-moving supersymmetry is preserved. The level-matching conditions \eqref{Eq:consistency} are satisfied.
The invariant lattice is
\begin{align}
    I=\{(p_L,P_L;0)|p_L\in\Lambda_R(D_4),P_L\in\Lambda_R(E_8\times E_8)\},
\end{align}
where an element of $\Lambda_R(E_8)$ lattice is written as $(n_1,\cdots,n_8)$ or $(n_1+1/2,\cdots,n_8+1/2)$ with $\sum n_i\in2\mathbb{Z}, n_i\in\mathbb{Z}$.
The dual lattice is
\begin{align}
    I^*=\{(p_L^*,P_L;0)|p_L\in\Lambda_W(D_4),P_L\in\Lambda_R(E_8\times E_8)\}.
\end{align}

\begin{itemize}
    \item Untwisted sector

    The choice of the shift vector~\eqref{eq:model2} breaks each $E_8$ to $E_7\times SU(2)$. Moreover, the group $SO(8)$ appears as we take $\Gamma^{4,4}(D_4)$ lattice.
    Therefore, the gauge group of the model is
    \begin{align}
        &E_7\times SU(2)\times E_7 \times SU(2)\times SO(8),
    \label{Eq:Hetero_gauge}\end{align}
    and $V=133+3+133+3+28=300$.
    The number of the tensor multiplet is one.
    In the untwisted sector, there are $224$ hypermultiplets. The charges are
    \begin{align}
        &(\mathbf{56},\mathbf{2},\mathbf{1},\mathbf{1},\mathbf{1})+(\mathbf{1},\mathbf{1},\mathbf{56},\mathbf{2},\mathbf{1}),
    \label{eq:model2_charge1}\end{align}
    under the group Eq.~\eqref{Eq:Hetero_gauge}. This can be understood that $\mathbf{248}$ of $E_8$ is broken to $(\mathbf{1},\mathbf{3})+(\mathbf{56},\mathbf{2})+(\mathbf{133},\mathbf{1})$ of $E_7\times SU(2)$.

    \item Twisted sector

    In the twisted sector, the degeneracy factor is $2$.
    The right-mover massless states are $(\mathbf{2},\mathbf{1})+2(\mathbf{1},\mathbf{1})$ under $SU(2)\times SU(2)$ little group. For the left-mover, the momentum lattice is shifted as $I^*+V_L$.
    The left-moving massless states are given by the states satisfying
    \begin{align}
        \frac{p_L^{*2}}{2}+\frac{(P_L+V_L)^2}{2}-1=0,
    \end{align}
    where $p_L^*\in\Lambda_W(D_4)$ and $P_L\in\Lambda_R(E_8\times E_8)$. It turns out that there are 224 states with $p_L^2=0$ and $(P_L+V_L)^2=2$, whose charge is
    \begin{align}
        (\mathbf{56},\mathbf{1},\mathbf{1},\mathbf{2},\mathbf{1})+(\mathbf{1},\mathbf{2},\mathbf{56},\mathbf{1},\mathbf{1})
    \label{eq:model2_charge2}\end{align}
    under the gauge group \eqref{Eq:Hetero_gauge}.
    Moreover, there are additional massless states with $p_L^2=1$ and $(P_L+V_L)^2=1$. The charge of these states is
    \begin{align}
        (\mathbf{1},\mathbf{2},\mathbf{1},\mathbf{2},\mathbf{8}_v)
        +(\mathbf{1},\mathbf{2},\mathbf{1},\mathbf{2},\mathbf{8}_s)
        +(\mathbf{1},\mathbf{2},\mathbf{1},\mathbf{2},\mathbf{8}_c)
    \label{eq:model2_charge3}\end{align}
    under the group \eqref{Eq:Hetero_gauge}. 
    In total, the number of hypermultiplets in the twisted sector is $320$. 

    \item Full spectrum

    By combining the untwisted and twisted sectors, the massless spectrum is
    \begin{align}
        G+T+300V+544H_c.
    \end{align}
    \end{itemize}

    The full spectrum is consistent with the gravitational anomaly cancellation. Next check is the condition $0=B_{Adj}-\sum n_R^i B_R^i$ ($\mathrm{Tr}F^4$ anomaly).
    This condition is trivially satisfied for $E_7$ and $SU(2)$ as $B_R=0$ for all the representations. 
    For $SO(8)$, this condition is nontrivially satisfied by using $B_{Adj}=0$, $B_{\mathbf{8}_v}=4$, and $B_{\mathbf{8}_s}=B_{\mathbf{8}_c}=-2$.
    The intersection matrix is
    \begin{align}
        \begin{pmatrix}
            8 & 2 & 26 & 2 & 26 & 2 \\
            2 & 0 & 12 & 0 & 12 & 0 \\
            26 & 12 & 24 & 12 & 24 & 12\\
            2 & 0 & 12 & 0 & 12 & 0 \\
            26 & 12 & 24 & 12 & 24 & 12\\
            2 & 0 & 12 & 0 & 12 & 0
        \end{pmatrix}
    \end{align}
    under $E_7\times SU(2)\times E_7\times SU(2)\times SO(8)$.
    All the anomaly can be satisfied by choosing $a=(-3,1), \,b_{E_7}=b_{E_7^\prime}=(1,-1)$, $ b_{SU(2)}=b_{SU(2)^\prime}=(7,5), \,b_{SO(8)}=(1,-1)$ with $j=-a$ for example.
    Finally, we show that the Kodaira condition, 
    \begin{align}
        j\cdot (-12a-9b_{E_7}-9b_{E_7^\prime}-6b_{SO(8)}-2b_{SU(2)}-2b_{SU(2)^\prime})>0
    \end{align}
    is not satisfied in this model.
    We parametrize $j$ as $c_1 b_{E_7}+c_2 b_{SU(2)}$ without loss of generality. Then, we obtain the condition
    \begin{align}
        -24(c_1+3c_2)>0.
    \end{align}
    However, this is incompatible with the positivity of the gauge kinetic term:
    \begin{align}
        &j\cdot b_{E_7}>0
        &&\text{and}
        &&j\cdot b_{SU(2)}>0
        &&\Leftrightarrow
        &&c_2>0
        &&\text{and}
        &&c_1+2c_2>0.
    \end{align}

Note that this result is independent of the choice of vectors as the choice of $-a$ fixes the other vectors. Similarly, it can be shown in the case of an even lattice with $-a=(2,2)$. Another choice is $-a=(4,1)$ but this  contradicts the fact that $-a$ is a characteristic class of the lattice \cite{Monnier:2018nfs}. Although, even in this case the solutions would be fixed and also not satisfying the Kodaira condition.

\subsubsection*{Heterotic model 3}

We again consider the heterotic string theory with the momentum lattice
\begin{align}
    \Gamma^{4,4}(A_2\oplus A_2)+2\Gamma^{8,0}(E_8).
\end{align}
The twist and shift are
\begin{align}
    &\phi_L=(0,0),
    \quad\quad\quad\quad\quad\quad\phi_R=\left(\frac{2}{3},\frac{2}{3}\right),
    \nonumber\\
    &V_L=\frac{1}{3}(1^6,0^2;0^8).
\label{eq:model3}\end{align}
The level-matching condition~\eqref{Eq:level_matching} is satisfied. The invariant lattice is
\begin{align}
    I=\{(p_L,P_L;0)|p_L\in\Lambda_R(A_2\oplus A_2),P_L\in\Lambda_R(E_8\times E_8)\},
\end{align}
and the dual lattice is
\begin{align}
    I^*=\{(p_L^*,P_L;0)|p_L\in\Lambda_W(A_2\oplus A_2),P_L\in\Lambda_R(E_8\times E_8)\}.
\end{align}

\begin{itemize}
    \item Untwisted sector

    The shift vector in Eq.~\eqref{eq:model3} breaks one of $E_8$ to $E_6\times SU(3)$. Therefore, the gauge group is
    \begin{align}
        E_6\times SU(3)\times E_8 \times SU(3)^2.
    \end{align}
    There are $81$ hypermultiplets charged under the gauge group:
    \begin{align}
        (\mathbf{27},\mathbf{3},\mathbf{1},\mathbf{1},\mathbf{1}).
    \label{eq:model3_charge1}\end{align}
    \item Twisted sector

    There are two twisted sectors. The degeneracy factor of each sector is $1$. The right-mover massless states are $(\mathbf{2},\mathbf{1})+2(\mathbf{1},\mathbf{1})$ under $SU(2)\times SU(2)$ little group. 
    The left-moving massless states are given by the states satisfying
    \begin{align}
        \frac{p_L^{*2}}{2}+\frac{(P_L\pm V_L)^2}{2}-1=0,
    \end{align}
    where $p_L^*\in\Lambda_W(A_2\oplus A_2)$, $P_L\in\Lambda_R(E_8\times E_8)$, and $\pm$ corresponds to the first and second twisted sectors. There are three types of massless hypermultiplets. First is the state with $p_L^{*2}/2=0$ and $(P_L\pm V_L)^2/2=1$. The charge of these states is
    \begin{align}
        (\mathbf{27},\mathbf{3},\mathbf{1},\mathbf{1},\mathbf{1}).
    \label{eq:model3_charge2}\end{align}
    The second type is the states with $p_L^{*2}/2=1/3$ and $(P_L\pm V_L)^2/2=2/3$. The charge is
    \begin{align}
        (\mathbf{27},\mathbf{1},\mathbf{1},\underline{\mathbf{3},\mathbf{1}}
        )+(\mathbf{27},\mathbf{1},\mathbf{1},\underline{\overline{\mathbf{3}},\mathbf{1}}
        ).
    \label{eq:model3_charge3}\end{align}
    The third type is the states with $p_L^{*2}/2=2/3$ and $(P_L\pm V_L)^2/2=1/3$. The charge is
    \begin{align}
        (\mathbf{1},\mathbf{3},\mathbf{1},\mathbf{3},\mathbf{3})+(\mathbf{1},\mathbf{3},\mathbf{1},\underline{\mathbf{3},\overline{\mathbf{3}}})+(\mathbf{1},\mathbf{3},\mathbf{1},\overline{\mathbf{3}},\overline{\mathbf{3}}).
    \label{eq:model3_charge4}\end{align}
    \item Full spectrum

    By combining the untwisted and twisted sectors, the massless spectrum is
    \begin{align}
        G+T+350V+594H_c.
    \end{align}
    This is consistent with the gravitational anomaly cancellation. The $\mathrm{Tr}F^4$ anomaly cancellation condition is trivially satisfied. 
\end{itemize}
The intersection matrix is
\begin{align}
    \begin{pmatrix}
        8 & 14 & 14 & -10 & 14 & 14\\
        14 & 12 & 12 & 0 & 12 & 12  \\
        14 & 12 & 12 & 0 & 12 & 12\\
        -10 & 0 & 0 & -12 & 0 & 0 \\
        14 & 12 & 12 & 0 & 12 & 12\\
        14 & 12 & 12 & 0 & 12 & 12
    \end{pmatrix}
\end{align}
under $E_6\times SU(3)\times E_8\times SU(3)^2$.
All the anomaly can be satisfied by choosing $a=(-3,1), \,b_{E_6}=b_{SU(3)}=b_{SU(3)^\prime}=b_{SU(3)^{\prime\prime}}=(4,2)$, $ b_{E_8}=(-2,-4)$ with $j=-a-b_{E_8}$ for instance.
The Kodaira condition, 
\begin{align}
        j\cdot (-12a-10b_{E_8}-8b_{E_6}-3b_{SU(3)}-3b_{SU(3)^\prime}-3b_{SU(3)^{\prime\prime}})>0
\end{align}
is not satisfied. To see this, we parametrize $j$ as $j=c_1 b_{E_6}+ c_2 b_{E_8}$. Then, the Kodaira condition is
\begin{align}
    c_1<0.
\end{align}
However, this contradicts the positivity of $SU(3)$ gauge kinetic term. Similarly, for the even lattice choice $-a=(2,2)$.

In fact, models $2,3$ where shown in \autoref{sec:6d} to be dual to F-theory models with bases giving rise to even string charge lattices. Since Higgsing does not affect the tensors since there is no such coupling we expect the lattice to remain the same.

\subsubsection*{Heterotic model 4}

We start from $E_8\times E_8$ string. We take $\Gamma^{4,4}(D_4)$ lattice with the twist (inspired by the CHL string)
\begin{align}
    &E_8\leftrightarrow E_8,
    &&\phi_R=\left(\frac{1}{2},\frac{1}{2}\right).
\end{align}
The gauge group is
\begin{align}
    E_8 \times SO(8).
\end{align}
$V=248+28=276$. The number of the hyper should be $H=520$ (Note that $T=1$).
The invariant lattice is
 \begin{align}
     I=\{(a;b,b)|a\in\Lambda_R(D_4),b\in\Lambda_R(E_8)\}.
 \end{align}
 The dual lattice is
 \begin{align}
     I^*=\left\{\left(\tilde{a};\frac{1}{2}(\tilde{b},\tilde{b})\right)|\tilde{a}\in\Lambda_W(D_4),\tilde{b}\in\Lambda_R(E_8)\right\}.
 \end{align}
There are $248$ untwisted hypers. Before orbifolding, there are $E_8\times E_8$ gauge bosons. By the CHL twist~\cite{Chaudhuri:1995fk,Chaudhuri:1995bf}, this becomes $E_8$ gauge bosons and $E_8$ adjoint hyper.
Therefore, the charges are
\begin{align}
    (\mathbf{248},\mathbf{1}).
\label{eq:model4_charge1}\end{align}
In the twisted sector, the mass formula for the left-mover is
\begin{align}
    &E_L=-1+\frac{1}{2}+N_B+\frac{p_L^2}{2},
    &&p_L\in I^*,
\end{align}
The left-moving massless states are
\begin{align}
    &24\text{ states with }p_L=(p_{so(8)};0,0), && p_{so(8)}\text{: weight vector of $\mathbf{8}_v$, $\mathbf{8}_s$, and $\mathbf{8}_c$,}\\
    &8\text{ states with }N_B=\frac{1}{2},\\
    &240\text{ states with }p_L=\left(0;\frac{1}{2}(p_{E_8},p_{E_8})\right),
    &&p_{E_8}\text{: root vector of $E_8$.}
\end{align}
These become hypermultiplets by combined with right-moving states.
The second and third states form the adjoint representation of $E_8$.
There are $272$ hypers in the twisted sector. The charges are
\begin{align}
    (\mathbf{1},\mathbf{8}_v)+(\mathbf{1},\mathbf{8}_s)+(\mathbf{1},\mathbf{8}_c)
    +(\mathbf{248},\mathbf{1}).
\label{eq:model4_charge2}\end{align}
The full spectrum is
\begin{align}
    G+T+276V+520H_c.
\end{align}
The intersection matrix is
\begin{align}
    \begin{pmatrix}
        8 & 10 & 1 \\
        10 & 12 & 0\\
        1 & 0 & -3
    \end{pmatrix},
\end{align}
under $E_8\times SO(8)$.
All the anomaly can be satisfied by choosing $a=(-3,1), \,b_{E_8}=(4,-2)$, $ b_{SO(8)}=(1,-2)$ with $j=-a$ for instance.

The Kodaira condition, 
\begin{align}
        j\cdot (-12a-10b_{E_8}-6b_{SO(8)})>0
\end{align}
is not satisfied. By parametrizing $j$ as $j=c_1 b_{E_8}+c_2 b_{SO(8)}$, the Kodaira condition becomes
\begin{align}
    c_2>0.
\end{align}
However, this contradicts  the positivity of the $SO(8)$ gauge kinetic term.

\subsection{Maximal Higgsing}\label{ap:maxhiggsing}
In this Section, we study the spectrum of our four models after the maximal Higgsing.
See \cite{Bershadsky:1996nh} for the chain of the Higgsing in 6d F-theory and the heterotic string.

\subsubsection*{Type II model 1} 
We start from the model 1. The gauge group is $U(1)^{12}$, and the charge of the hypermultiplets is given in Eq.~\eqref{eq:model1_charge}.
By turning on the VEV to $\left(\underline{\pm1,0,0,0},0^8\right)$ hypermultiplets, $U(1)^4$ are Higgsed.
The spectrum after the Higgsing is 20 neutral hypermultiplets with the gauge group $U(1)^8$.

The model after the Higgsing can also be described in terms of an orbifold. The lattice is $\Gamma^{4,4}(D_4)$ with a $\mathbb{Z}_4$ twist
\begin{align}
    &\phi_L=\left(\frac{1}{4},\frac{3}{4}\right),
    &&\phi_R=\left(\frac{1}{4},\frac{1}{4}\right).
\end{align}
The spectrum is $G+T+2V+2H_0$ in the untwisted sector, $8T+8H_0$ in the $g$ and $g^3$-twisted sectors, and $6V+10H_0$ in the $g^2$-twisted sector.
The full massless spectrum is $G+9T+8V+20H_0$.

\subsubsection*{Heterotic model 2}
The gauge group is $[E_7\times SU(2)]^2\times SO(8)$, and the charge of the hypermultiplets is Eqs.~\eqref{eq:model2_charge1}\eqref{eq:model2_charge2}\eqref{eq:model2_charge3}. In the following, we show that the whole gauge group can be Higgsed.

First we note that
\begin{align}
    &E_7\times SU(2) &&\to &&E_6\times U(1)\times U(1)\nonumber\\
    &(\mathbf{56},\mathbf{2})
    &&\to
    &&\overbrace{(\mathbf{27},-1,-1)
    +(\overline{\mathbf{27}},1,1)}^{\text{massive}}+(\mathbf{27},-1,1)+(\overline{\mathbf{27}},1,-1)
    \nonumber\\
    & && &&+(\mathbf{1},3,1)+\underbrace{(\mathbf{1},3,-1)+(\mathbf{1},-3,1)+(\mathbf{1},-3,-1)}_{\text{massive}}.
\end{align}
This means that by turning on the VEV to $(\mathbf{1},3,1)$, $E_7\times SU(2)$ is broken to $E_6\times U(1)^\prime$, where $U(1)^\prime$ is a linear combination of two $U(1)$ under which $(\mathbf{1},3,1)$ is neutral. At the same time, $57$ hypermultiplets become massive to form massive vector multiplets.

The spectrum after Higgsing is
\begin{align}
    &E_7\times SU(2) &&
    \to
    &&E_6\times U(1)^\prime\nonumber\\
    &(\mathbf{56},\mathbf{2})
    &&\to 
    &&(\mathbf{27},-4)+(\overline{\mathbf{27}},4)+(\mathbf{1},0).
\end{align}
This Higgsing can be done for each $E_7\times SU(2)$ factor, and so the full gauge group is Higgsed to
\begin{align}
    &[E_7\times SU(2)]^2\times SO(8) &&\to &&[E_6\times U(1)^\prime]^2\times SO(8).
\end{align} 
Under the Higgsing, one of the hypermultiplets decompose as
\begin{align}
    &(\mathbf{56},\mathbf{1},\mathbf{1},\mathbf{2},\mathbf{1})
    &&\to
    &&(\mathbf{27},-1,\mathbf{1},\pm3,\mathbf{1})+(\overline{\mathbf{27}},1,\mathbf{1},\pm3,\mathbf{1})\nonumber\\
    & && &&+(\mathbf{1},3,\mathbf{1},\pm3,\mathbf{1})
    +(\mathbf{1},-3,\mathbf{1},\pm3,\mathbf{1})
\end{align}
The gauge group $U(1)^{\prime2}$ can be Higgsed by turning on the VEV to $(\mathbf{1},3,\mathbf{1},\pm3,\mathbf{1})$.

At this stage, the gauge group is $E_6^2\times SO(8)$, and the hypermultiplets are
\begin{align}
    3(\underline{\mathbf{27},\mathbf{1}},\mathbf{1})+3(\underline{\overline{\mathbf{27}},\mathbf{1}},\mathbf{1})+4(\mathbf{1},\mathbf{1},\mathbf{8}_v)+4(\mathbf{1},\mathbf{1},\mathbf{8}_s)+4(\mathbf{1},\mathbf{1},\mathbf{8}_c)+10(\mathbf{1},\mathbf{1},\mathbf{1}).
\end{align}

In order to Higgs $E_6$ completely, $3(\mathbf{27}+\overline{\mathbf{27}})$ is enough:
\begin{align}
    &E_6 &&\xrightarrow[]{\mathbf{27}+\overline{\mathbf{27}}}
    && SO(10)
    &&\xrightarrow[]{\mathbf{16}+\overline{\mathbf{16}}}
    && SU(5)
    \nonumber\\
    &3(\mathbf{27}+\overline{\mathbf{27}})
    &&\xrightarrow[]{\phantom{\mathbf{27}+\overline{\mathbf{27}}}}
    && \overbrace{(\mathbf{16}+\overline{\mathbf{16}}+\mathbf{1})}^{\text{massive}}+2(\mathbf{16}+\overline{\mathbf{16}})
    &&\xrightarrow[]{\phantom{\mathbf{16}+\overline{\mathbf{16}}}}
    && 
    \overbrace{(\mathbf{10}+\overline{\mathbf{10}}+\mathbf{1})}^{\text{massive}}
    \nonumber\\
    & && &&+6(\mathbf{10})+5(\mathbf{1})
    && &&+(\mathbf{10}+\overline{\mathbf{10}})
    \nonumber\\
    & && && && &&
    +8(\mathbf{5}+\overline{\mathbf{5}})+8(\mathbf{1})
\end{align}

\begin{align}
    &\xrightarrow[]{\mathbf{5}+\overline{\mathbf{5}}}
    &&SU(4)&&\xrightarrow[]{\mathbf{4}+\overline{\mathbf{4}}}
    &&SU(3)&&\xrightarrow[]{\mathbf{3}+\overline{\mathbf{3}}}
    &&SU(2)
    \nonumber\\
    &\xrightarrow[]{\phantom{\mathbf{5}+\overline{\mathbf{5}}}}
    && \overbrace{(\mathbf{4}+\overline{\mathbf{4}}+\mathbf{1})}^{\text{massive}}+(\mathbf{6}+\overline{\mathbf{6}})
    &&\xrightarrow[]{\phantom{\mathbf{4}+\overline{\mathbf{4}}}}
    &&\overbrace{(\mathbf{3}+\overline{\mathbf{3}}+\mathbf{1})}^{\text{massive}}
    &&\xrightarrow[]{\phantom{\mathbf{3}+\overline{\mathbf{3}}}}
    &&\overbrace{2(\mathbf{2})+(\mathbf{1})}^{\text{massive}}
    \nonumber\\
    &
    && +8(\mathbf{4}+\overline{\mathbf{4}})+23(\mathbf{1})
    && &&+9(\mathbf{3}+\overline{\mathbf{3}})+38(\mathbf{1}) &&  && 16(\mathbf{2})+55(\mathbf{1}) 
    \nonumber\\
    \nonumber\\
    &\xrightarrow[]{2(\mathbf{2})}
    &&\text{Nothing}
    \nonumber\\
    &\xrightarrow[]{\phantom{2(\mathbf{2})}}
    &&\overbrace{3(\mathbf{1})}^{\text{massive}}+84(\mathbf{1})
\end{align}

Similarly, $4(\mathbf{8}_v+\mathbf{8}_s+\mathbf{8}_c)$ can Higgs $SO(8)$ completely.

\subsubsection*{Heterotic model 3}
The gauge group is $E_6\times SU(3)\times E_8 \times SU(3)^2$. The representation of the hypermultiplets is Eqs.~\eqref{eq:model3_charge1}\eqref{eq:model3_charge2}\eqref{eq:model3_charge3}\eqref{eq:model3_charge4}. We show that $E_8$ is the only gauge group that remains after the maximal Higgsing.

By turning on the VEV to $2(\mathbf{27},\mathbf{3},\mathbf{1},\mathbf{1},\mathbf{1})$, $E_6\times SU(3)$ is broken to $SO(10)\times SU(2)\times U(1)$:
\begin{align}
    &E_6\times SU(3)
    &&\to
    &&SO(10)\times SU(2)\times U(1)
    \nonumber\\
    &2(\mathbf{27},\mathbf{3})
    &&\to
    &&\underbrace{2(\mathbf{16},\mathbf{1},3)+2(\mathbf{1},\mathbf{2},3)+(\mathbf{1},\mathbf{1},0)}_{\text{massive}}
    \nonumber\\
    & && && +2(\mathbf{16},\mathbf{2},-3)
    +2(\mathbf{10},\mathbf{2},0)
    +2(\mathbf{10},\mathbf{1},6)
    +(\mathbf{1},\mathbf{1},0).
\end{align}
Moreover, the VEV of $2(\mathbf{10},\mathbf{2},0)$ breaks $SO(10)\times SU(2)\times U(1)$ to $SO(9)\times U(1)$:
\begin{align}
    &SO(10)\times SU(2)\times U(1)
    &&\to
    &&SO(9)\times U(1)
    \nonumber\\
    &2(\mathbf{10},\mathbf{2},0)+2(\mathbf{16},\mathbf{2},-3)
    &&\to
    &&\overbrace{(\mathbf{9},0)+3(\mathbf{1},0)}^{\text{massive}}+3(\mathbf{9},0)+(\mathbf{1},0)
    \nonumber\\
    &+2(\mathbf{10},\mathbf{1},6)
    +(\mathbf{1},\mathbf{1},0)
    &&
    &&+4(\mathbf{16},-3)+2(\mathbf{9},6)+2(\mathbf{1},6)+(\mathbf{1},0).
\end{align}
Next, $2(\mathbf{1},6)$ can break $U(1)$ completely.
Combining these steps, we observe
\begin{align}
    &E_6\times SU(3)
    &&\to
    && SO(9)
    \nonumber\\
    &2(\mathbf{27},\mathbf{3})
    &&\to
    &&4(\mathbf{16})+5(\mathbf{9})+3(\mathbf{1}).
\end{align}
By using $3(\mathbf{9})$, we can Higgs $SO(9)$ as $SO(9)\to SO(8)\to SO(7)\to SU(4)$:
\begin{align}
    &SO(9)
    &&\xrightarrow[]{3(\mathbf{9})}
    &&SU(4)
    \nonumber\\
    &4(\mathbf{16})+5(\mathbf{9})+3(\mathbf{1})
    &&\xrightarrow[]{\phantom{3(\mathbf{9})}}
    &&16(\mathbf{4})+2(\mathbf{6})+15(\mathbf{1}).
\end{align}
In order to Higgs $SU(4)$ completely, $16(\mathbf{4})$ is enough. 

At this stage, the massless spectrum is
\begin{align}
    76(\mathbf{1},\mathbf{1},\mathbf{1})+54(\mathbf{1},\underline{\mathbf{3},\mathbf{1}})
    +6(\mathbf{1},\mathbf{3},\mathbf{3})
    +3(\mathbf{1},\underline{\mathbf{3},\overline{\mathbf{3}}})
\end{align}
under the gauge group $E_8\times SU(3)^2$.
Clearly, $54(\mathbf{1},\underline{\mathbf{3},\mathbf{1}})$ is enough to Higgs $SU(3)^2$. Consequently, after the maximal Higgsing, there are neutral hypermultiplets with the gauge group $E_8$.

\subsubsection*{Heterotic model 4}
The gauge group is $E_8\times SO(8)$, and the charge of hypermultiplets is Eqs.~\eqref{eq:model4_charge1}\eqref{eq:model4_charge2}.
We show that $SU(3)$ gauge group remains after the maximal Higgsing.

The $E_8$ gauge group is completely Higgsed by two adjoint representation. The 248 neutral hypermultiplets are in the spectrum after the Higgsing. As for $SO(8)$ part, the $SU(3)$ gauge group remains even after the maximal Higgsing:
\begin{align}
    &SO(8) && \xrightarrow[]{\mathbf{8}_v} && SO(7) && \xrightarrow[]{\mathbf{8}} && G_2 && \xrightarrow[]{\mathbf{7}} && SU(3)
    \nonumber\\
    &(\mathbf{8}) +(\mathbf{8}_s)
    &&\xrightarrow[]{\phantom{\mathbf{8}_v}}
    &&\overbrace{(\mathbf{7})}^{\text{massive}}+(\mathbf{1})
    &&\to && \overbrace{(\mathbf{7})}^{\text{massive}}+(\mathbf{7}) 
    && \to && 
    \overbrace{(\mathbf{3}+\overline{\mathbf{3}})}^{\text{massive}}
    \nonumber\\
    &+(\mathbf{8}_c)
    &&
    &&+2(\mathbf{8})
    && && +3(\mathbf{1})
    && && +4(\mathbf{1})
\end{align}
Therefore, the spectrum after the Higgsing is $SU(3)$ with $252$ singlets.

\section{Details of the 5d models}\label{ap:5dmodels}
As described in \autoref{sec:5d} for the type II models, we choose  the momentum lattice
\begin{align}
    \Gamma^{5,5}=\Gamma^{4,4}+\Gamma^{1,1}.
\end{align}
and for the heterotic models, we choose 
\begin{align}
    \Gamma^{21,5}=\Gamma^{20,4}+\Gamma^{1,1}.
\end{align}

A choice of discrete twist  $\mathbb{Z}_N$ needs to specified that acts as a twist for $\Gamma^{4,4}$ and $\Gamma^{20,4}$, while it acts as a shift for $\Gamma^{1,1}$. Thanks to this shift in $\Gamma^{1,1}$, the states in the twisted sector become massive for a sufficiently large radius as reviewed in  (see Appendix~\ref{sec:shift}). This means that only the untwisted sectors are important for this discussion.

\textbf{Untwisted Sector}

Suppose that the $\mathbb{Z}_N$ twist on $\Gamma^{4,4}$ or $\Gamma^{20,4}$ breaks half of right-supersymmetry and all (if any) of left-supersymmetry. We denote the number of zero eigenvalues of the twist by $z_L$ and $z_R$ (counted in real) for a left and right twist, respectively. Note that $z_R=0$ while $z_L$ takes values from 0 to 4 for type II and from 0 to 20 for the heterotic case. Consequently, the untwisted spectrum is as follows:
\begin{align}
    [\mathbf{3}_v+(z_L+1)(\text{scalar})]\times
    [\mathbf{3}_v+(\text{scalar})+2(\text{spinor})]
    +\cdots,
\label{Eq:5d_untwisted}\end{align}
where $\mathbf{3}_v$ represents the 5d vector, and $(\text{scalar})$ indicates a real scalar.
The symbol $\cdots$ denotes other scalars, spinors, and vectors which potentially appear corresponding to states where the left and right phases cancel out nontrivially.
For such additional states in the NSNS sector, we obtain other hypermultiplets. For additional states in the RR sector, we obtain other vector multiplets. 

The bosonic spectrum is
\begin{align}
    \eqref{Eq:5d_untwisted}=
    (\mathbf{5}_{g_{\mu\nu}}+\mathbf{3}_v)
    +(z_L+2)\mathbf{3}_v
    +(z_L+2)(\text{scalar})
    +\cdots,
\label{eq:5d_spectrum}\end{align}
where $\mathbf{5}_{g_{\mu\nu}}$ represents the 5d graviton, and $\mathbf{5}_{g_{\mu\nu}}+\mathbf{3}_v$ is the bosonic part of the 5d gravity multiplet. 
Based on the discussion above, we observe
\begin{itemize}
    \item We must not have extra states in the NSNS sector, where the phases in the left and right-mover nontrivially cancel, otherwise we will get hypermultiplets.\footnote{This is why symmetric orbifolds do not lead to models without hypermultiplets.}
    \item From Eq.~\eqref{eq:5d_spectrum}, we have $z_L+2$ vector multiplets. This is the number of vectors for the heterotic, while other vector multiplets may appear from the RR sector for type II string.
    \item When the determinant of the left twist is one, then the number of the vector multiplets obtained in this way is always even (this is also observed in \cite{Gkountoumis:2023fym}). This is because nonzero eigenvalues appear in complex pairs. 
\end{itemize}

We provide a detailed explanation of the twist $\Gamma^8(E_8)\leftrightarrow\Gamma^8(E_8)$ in Table~\ref{Tab:Freely_acting}. This is the CHL twist~\cite{Chaudhuri:1995fk}, which exchanges two $E_8$s.
 The invariant lattice and its dual are
 \begin{align}
     &I_{CHL}=\{(a,a)|a\in\Lambda_R(E_8)\},
     &&I_{CHL}^*=\left\{\frac{1}{2}(b,b)|b\in\Lambda_R(E_8)\right\}.
 \end{align}
 The twist $\Gamma^{2,2}(A_2)\leftrightarrow\Gamma^{2,2}(A_2)$ is similar.

\section{Other constructions}\label{sec:DJK}

In this appendix we comment on earlier studies of the 4d $\mathcal{N}=2$ string landscape with a small number of moduli using other techniques.
\begin{itemize}
    \item Based on the free fermionic construction~\cite{Kawai:1986va,Lerche:1986cx,Kawai:1986ah,Antoniadis:1986rn}, a hyper-free 4d $\mathcal{N}=2$ models with one vector multiplet is constructed in \cite{Dolivet:2007sz}. However, contrary to the claim, we find the model leads to 4d $\mathcal{N}=3$ supersymmetry. The details are shown in the next subsection. 
    \item The model with effective Hodge numbers $(h_{11},h_{21})=(1,1)$ is constructed in \cite{Kiritsis:2008mu,Anastasopoulos:2009kj}. By using the Gepner model~\cite{Israel:2013wwa} or the duality twist~\cite{Hull:2017llx}, the hyper-free model with 3 vector multiplets is constructed, which is consistent with our result.
\end{itemize}
Interestingly, the minimal matter content seems to be $H+V=3$ (except for 3d models in Section~\ref{sec:3d_models}), where $H$ and $V$ is the number of hypermultiplets and vector multiplets respectively.

\subsection{Check of the DJK model}

This model is claimed to possess $\mathcal N=2$ supersymmetry and no hypers in its spectrum. We show that there is $\mathcal N=3$ supersymmetry, contrary to the original claim.

\subsection*{Setup}
The DJK model performs a generalized GSO projection using
\begin{align}
    b_1 &= \{\psi^\mu,\chi^{1,2},y^{3,4},y^{5,6},y^1,w^1\mid \bar y^5,\bar w^5\},\\
    \bar b_1 &= \{\bar \psi^\mu,\bar\chi^{1,2},\bar y^{3,4},\bar y^{5,6},\bar y^1,\bar w^1\mid  y^5, w^5\},\\
    \bar b_2 &= \{\bar \psi^\mu,\bar\chi^{3,4},\bar y^{1,2},\bar w^{5,6},\bar y^3,\bar w^3\mid  y^6, w^6\},\\
    \bar b_3 &= \{\bar \psi^\mu,\bar\chi^{5,6},\bar w^{1,2},\bar w^{3,4},\bar y^6,\bar w^6\mid  y^2, w^2\},
\end{align}
in addition to $F,S,\bar S$ (see \cite{Antoniadis:1986rn} for the definitions). The $y^i,w^i$ are fermionizations of the internal coordinates
\begin{align}
    \normord{e^{i X_L^i}} &= \frac 1 {\sqrt{2}}(y^i+iw^i),\\
    \normord{e^{i X_R^i}} &= \frac 1 {\sqrt{2}}(\bar y^i+i\bar w^i).
\end{align}
The correspondence between generalized GSO actions and bosonic actions are
\begin{align}\label{eq:bosonic-yi}
    y^i &: \quad y^i+iw^i \mapsto -y^i +iw^i & \iff e^{iX_L^i}\mapsto -e^{-iX_L^i} &\iff X_L^i \mapsto -X_L^i +\pi,\\
    w^i &: \quad y^i+iw^i \mapsto y^i -iw^i &\iff e^{iX_L^i}\mapsto e^{-iX_L^i} &\iff X_L^i \mapsto -X_L^i,\\\label{eq:bosonic-yiwi}
    y^iw^i &: \quad y^i+iw^i \mapsto -y^i -iw^i &\iff e^{iX_L^i}\mapsto -e^{iX_L^i} &\iff X_L^i \mapsto X_L^i +\pi,
\end{align}
and similarly for the right movers.

\subsubsection*{Basis change}
Performing the basis change as in \cite{Farragi:2023},
\begin{align}
    \tilde b_3 &= \bar b_1 +\bar b_2 +\bar S =\{\bar \psi^\mu, \bar \chi^{5,6}, \bar y^2, \bar w^1, \bar y^4, \bar w^3, \bar y^{5,6},\bar w^{5,6}\mid y^{5,6},w^{5,6}\}\\
    \bar B_3 &= \tilde b_3 + \bar b_3 = \{\bar y^{2,4,5}\bar w^{2,4,5}\mid y^{2,5,6}w^{2,5,6}\}.
\end{align}
We can use a pure shift $\bar B_3$ instead of $\bar b_3$ in the generators.

\subsection*{Supersymmetry}\label{sec:susy}
We find 2 gravitinos in the $S$ sector. 
However, we also find a gravitino in the $\bar S$ sector. In total, we get $4d$ $\mathcal N=3$ instead of the claimed $\mathcal N=2$.

The gravitino in the $\bar S$ sector have the form $\psi^\mu_{-\frac 1 2} \ket{0}\otimes \ket{\bar S}$. Here, $\ket{\bar S}$ denotes a spinor in 10d
\begin{align}\label{eq:spinor10d}
    \ket{\bar S} = \ket{s_1,s_2,s_3,s_4},
\end{align}
where $s_i=\pm$. Compactifying to 4d, we distinguish between the spacetime and internal components with a line.
\begin{align}
    \ket{s_1\mid s_2,s_3,s_4}.
\end{align}
The generator $\bar S$ halves the number of possibilities for $s_i$. In other words,
\begin{align}\label{eq:s1-4}
    (-1)^{\bar S}=-1\implies s_1s_2s_3s_4=-1.
\end{align}
We can also see that $(-1)^F=1$ and $(-1)^{S}=-1$ are satisfied by the gravitino due to the presence of the oscillator on the left $\psi^{\mu}_{-\frac 1 2}$.

In $4$ dimensions, we therefore have $4$ gravitinos
\begin{align}
    \ket{s_1 \mid s_2,s_3,-s_1 s_2 s_3},
\end{align}
corresponding to the sign choices of $s_2,s_3$. Then, the projections require
\begin{align}\label{eq:s12}
    (-1)^{\bar b_1} &=-1\implies s_1s_2=-1\\\label{eq:s13}
    (-1)^{\bar b_2} &= -1 \implies s_1 s_3=-1.
\end{align}
The generator $\bar b_3$ does not impose an additional condition (as expected since it can be replaced by a pure translation $\bar B_3$). Also, $b_1$ does not impose a condition since it is a pure translation on the right.

So $\bar b_1$ and $\bar b_2$ together reduce the number of gravitino by a quarter. As a result, we get exactly one gravitino in the $\bar S$ sector
\begin{align}
    \psi^\mu_{-\frac 1 2 }\ket{0}\otimes \ket{s_1 | -s_1, -s_1, -s_1}.
\end{align}
In summary, we used \eqref{eq:s1-4}, \eqref{eq:s12}, \eqref{eq:s13} on \eqref{eq:spinor10d}. 

Together with the two gravitino from the $S$ sector, we have $3$ gravitino in total, hence $\mathcal N=3$ supersymmetry.

\section{No 4d cyclic asymmetric orbifolds with one vector multiplet}\label{sec:no_4d_model}
Here we show that no 4d cyclic asymmetric orbifolds with one vector multiplet both for the heterotic and type II. An idea is to show that there is no shift vector which lifts all the twisted sectors. In other words, there exists $k$-th twisted sector where any shift vector $v$ belongs to $k v\in  I^*$ under the condition for the shift vector $N v\in I$ and $Nv^2/2\in \mathbb{Z}$. Here $N$ is the order of the orbifold. We assume that the twist is inner automorphism.

First, we note that there is no invariant lattice for the left-mover, and two eigenvalue one for the right-mover in order to realize the model with eight supercharges and one vector multiplet. Therefore, the invariant lattice is the right-mover root lattice of the Lie algebra. We consider the cases where the $I=\Lambda_R(D_6)$ and $\Lambda_R(E_6)$. The same argument holds for the other choices of the lattices.

\begin{itemize}
    \item $I=\Lambda_R(D_6)$

    The right twist must preserve 4d $\mathcal{N}=2$ supersymmetry, and the possible choices are
    \begin{align}
        &(\bar{1},\bar{1},\bar{1},\bar{1})\Rightarrow
        \phi_R=\left(\frac{1}{2},\frac{1}{2}\right),
        \\
        &(3,3)\Rightarrow
        \phi_R=\left(\frac{1}{3},\frac{1}{3}\right),
        \\
        &(2,2,\bar{1},\bar{1})\Rightarrow
        \phi_R=\left(\frac{1}{2},\frac{1}{2}\right),
        \\
        &(\bar{2},\bar{2})\Rightarrow
        \phi_R=\left(\frac{1}{4},\frac{1}{4}\right).
    \end{align}
    Note that the Weyl group of $D_n$ is given by the permutation and even number of sign flips of the orthonormal basis, $e_1,e_2,\cdots,e_n$.
    Here $``\,i\,"$ means the permutation $e_1\to e_2\to\cdots\to e_i\to e_1$, and $``\,\bar{i}\,"$ means the permutation with the sign flip, $e_1\to e_2\to\cdots\to e_i\to -e_1$.
    For each twist, the invariant lattice $I$ and its dual $I^*$ are given by
    \begin{align}
        (\bar{1},\bar{1},\bar{1},\bar{1}):\,&I=n_1(0,0,0,0,1,1)+n_2(0,0,0,0,1,-1),
        \nonumber\\
        &I^*=n_1\left(0,0,0,0,\frac{1}{2},\frac{1}{2}\right)+n_2\left(0,0,0,0,\frac{1}{2},-\frac{1}{2}\right),
        \\
        (3,3):\,&I=n_1(1,1,1,1,1,1)+n_2(1,1,1,-1,-1,-1),
        \nonumber\\
        &I^*=n_1\left(\frac{1}{6},\frac{1}{6},\frac{1}{6},\frac{1}{6},\frac{1}{6},\frac{1}{6}\right)+n_2\left(\frac{1}{6},\frac{1}{6},\frac{1}{6},-\frac{1}{6},-\frac{1}{6},-\frac{1}{6}\right),
        \\
        (2,2,\bar{1},\bar{1}):\,&I=n_1(1,1,0,0,0,0)+n_2(0,0,1,1,0,0),
        \nonumber\\
        &I^*=n_1\left(\frac{1}{2},\frac{1}{2},0,0,0,0\right)+n_2\left(0,0,\frac{1}{2},\frac{1}{2},0,0\right),
        \\
        (\bar{2},\bar{2}):\,&I=n_1(0,0,0,0,1,1)+n_2(0,0,0,0,1,-1),
        \nonumber\\
        &I^*=n_1\left(0,0,0,0,\frac{1}{2},\frac{1}{2}\right)+n_2\left(0,0,0,0,\frac{1}{2},-\frac{1}{2}\right)
    \end{align}
    in terms of the orthonormal basis.
    We can show that there are no vectors $v$ which satisfy both $N v\in I$ and $k v\notin I^*$ for $k=1,\cdots,N-1$.
    Let us consider the $(\bar{1},\bar{1},\bar{1},\bar{1})$ twist for  illustration purposes. It is easy to see that $Nv/2\in I^*$ is satisfied for any $v$. The same argument applies to the other cases.

    \item $I=\Lambda_R(E_6)$
    
    From Table~9 in \cite{Carter1970}, the twist preserving half supersymmetry is either $A_1^4$ (order $2$), $A_2^2$ (order $3$), or $D_4(a_1)$ (order $4$). 
    
    We start from the twist $A_2^2$. The $E_6$ has a $A_2^3$ as a maximal subgroup. The twist $A_2^2$ is the $2\pi/3$ rotation of the two $A_2$s.
    The invariant lattice is the root lattice of the third $A_2$, which is $I=n_1(1,-1,0)+n_2(0,1,-1)$. The dual of $I$ is $I^*=n_1(1/3,1/3,-2/3)+n_2(1/3,-2/3,1/3)$.
    Given the conditions $N v\in I$ and $Nv^2/2\in \mathbb{Z}$, we parametrize the shift vector as
    \begin{align}
        v=\left(\frac{m_1}{N},-\frac{m_1+m_2}{N},\frac{m_2}{N}\right),
    \end{align}
    where $m_{1,2}\in\mathbb{Z}$ and 
    \begin{align}
        \frac{m_1^2-m_1m_2+m_2^2}{N}\in\mathbb{Z}.
    \label{eq:A2_v_condition}\end{align}
    From the condition~\eqref{eq:A2_v_condition}, we obtain $m_1=m_2$ mod $3$ since $N$ is a multiple of $3$.
    Then, for $g^{N/3}$-twisted sector, $N v/3$ is an element of $I^*$:
    \begin{align}
        \frac{N}{3}v\cdot I= \frac{1}{3}\bigg((2m_1+m_2)n_1-(m_1+2m_2)n_2\bigg) \in \mathbb{Z}.
    \end{align}
    
    In order to deal with $A_1^4$ and $D_4(a_1)$ twists, we use the following parametrization of the $E_6$ root lattice~\cite{Solr-2759686}:
    \begin{align}
        &\Lambda_R(E_6)=n_1(1,1,0,0,0,0,0,0)+n_2(-1,1,0,0,0,0,0,0)
        \nonumber\\
        &+n_3(0,-1,1,0,0,0,0,0)+n_4(0,0,-1,1,0,0,0,0)+n_5(0,0,0,-1,1,0,0,0)
        \nonumber\\
        &+n_6\left(\frac{1}{2},-\frac{1}{2},-\frac{1}{2},\frac{1}{2},-\frac{1}{2},-\frac{1}{2},-\frac{1}{2},\frac{1}{2}\right)
    \end{align}
    in terms of an eight-dimensional orthonormal lattice.
    Here the first four entry corresponds to orthonormal coordinate of the $D_4$ lattice, and $D_4(a_1)$ is the $(\overline{2},\overline{2})$ twist for these four coordinates. The $A_1^4$ is a maximal subgroup of $D_4$, and the $A_1^4$ is the reflection for all $A_1$s. 
    The invariant lattice and its dual in these cases are
    \begin{align}
        &I=n_1(0,0,0,0,-1,-1,-1,1)+n_2(0,0,0,0,2,0,0,0),
        \nonumber\\
        &I^*=n_1\left(0,0,0,0,0,-\frac{1}{3},-\frac{1}{3},\frac{1}{3}\right)+n_2\left(0,0,0,0,\frac{1}{2},-\frac{1}{6},-\frac{1}{6},\frac{1}{6}\right).
    \end{align}
    We parametrize the shift vector as
    \begin{align}
        v=\left(0,0,0,0,\frac{2m_2-m_1}{N},-\frac{m_1}{N},-\frac{m_1}{N},\frac{m_1}{N}\right),
    \end{align}
    where $m_{1,2}\in\mathbb{Z}$. From $Nv^2/2=\mathbb{Z}$, we obtain
    \begin{align}
        N\frac{v^2}{2}=\frac{2(m_1^2-m_1m_2+m_2^2)}{N}\in\mathbb{Z}.
    \end{align}
    Then, in the $g^{N/2}$ twisted sector, we get
    \begin{align}
        \frac{N}{2}\cdot I=2m_1n_1-m_2n_1-m_1n_2+2m_2n_2\in\mathbb{Z}.
    \end{align}
\end{itemize}

\bibliographystyle{utphys} 
\bibliography{refs}

\providecommand{\href}[2]{#2}\begingroup\raggedright\begin{thebibliography}{10}

\bibitem{Vafa:2005ui}
C.~Vafa,  {\em {The String landscape and the swampland}}, \href{http://www.arXiv.org/abs/hep-th/0509212}{{\tt hep-th/0509212}}.

\bibitem{Palti:2019pca}
E.~Palti,  {\em {The Swampland: Introduction and Review}}, Fortsch. Phys. {\bf 67} (2019), no.~6, 1900037 [\href{http://www.arXiv.org/abs/1903.06239}{{\tt 1903.06239}}].

\bibitem{vanBeest:2021lhn}
M.~van Beest, J.~Calder\'on-Infante, D.~Mirfendereski and I.~Valenzuela,  {\em {Lectures on the Swampland Program in String Compactifications}}, Phys. Rept. {\bf 989} (2022) 1--50 [\href{http://www.arXiv.org/abs/2102.01111}{{\tt 2102.01111}}].

\bibitem{Grana:2021zvf}
M.~Gra\~na and A.~Herr\'aez,  {\em {The Swampland Conjectures: A Bridge from Quantum Gravity to Particle Physics}}, Universe {\bf 7} (2021), no.~8, 273 [\href{http://www.arXiv.org/abs/2107.00087}{{\tt 2107.00087}}].

\bibitem{Agmon:2022thq}
N.~B. Agmon, A.~Bedroya, M.~J. Kang and C.~Vafa,  {\em {Lectures on the string landscape and the Swampland}}, \href{http://www.arXiv.org/abs/2212.06187}{{\tt 2212.06187}}.

\bibitem{Kim:2019ths}
H.-C. Kim, H.-C. Tarazi and C.~Vafa,  {\em {Four-dimensional $\mathbf{\mathcal{N}=4}$ SYM theory and the swampland}}, Phys. Rev. D {\bf 102} (2020), no.~2, 026003 [\href{http://www.arXiv.org/abs/1912.06144}{{\tt 1912.06144}}].

\bibitem{Montero:2020icj}
M.~Montero and C.~Vafa,  {\em {Cobordism Conjecture, Anomalies, and the String Lamppost Principle}}, JHEP {\bf 01} (2021) 063 [\href{http://www.arXiv.org/abs/2008.11729}{{\tt 2008.11729}}].

\bibitem{Cvetic:2020kuw}
M.~Cveti\v{c}, M.~Dierigl, L.~Lin and H.~Y. Zhang,  {\em {String Universality and Non-Simply-Connected Gauge Groups in 8d}}, Phys. Rev. Lett. {\bf 125} (2020), no.~21, 211602 [\href{http://www.arXiv.org/abs/2008.10605}{{\tt 2008.10605}}].

\bibitem{Hamada:2021bbz}
Y.~Hamada and C.~Vafa,  {\em {8d supergravity, reconstruction of internal geometry and the Swampland}}, JHEP {\bf 06} (2021) 178 [\href{http://www.arXiv.org/abs/2104.05724}{{\tt 2104.05724}}].

\bibitem{Bedroya:2021fbu}
A.~Bedroya, Y.~Hamada, M.~Montero and C.~Vafa,  {\em {Compactness of brane moduli and the String Lamppost Principle in d \ensuremath{>} 6}}, JHEP {\bf 02} (2022) 082 [\href{http://www.arXiv.org/abs/2110.10157}{{\tt 2110.10157}}].

\bibitem{deBoer:2001wca}
J.~de~Boer, R.~Dijkgraaf, K.~Hori, A.~Keurentjes, J.~Morgan, D.~R. Morrison and S.~Sethi,  {\em {Triples, fluxes, and strings}}, Adv. Theor. Math. Phys. {\bf 4} (2002) 995--1186 [\href{http://www.arXiv.org/abs/hep-th/0103170}{{\tt hep-th/0103170}}].

\bibitem{Aharony:2007du}
O.~Aharony, Z.~Komargodski and A.~Patir,  {\em {The Moduli space and M(atrix) theory of 9d N=1 backgrounds of M/string theory}}, JHEP {\bf 05} (2007) 073 [\href{http://www.arXiv.org/abs/hep-th/0702195}{{\tt hep-th/0702195}}].

\bibitem{Font:2020rsk}
A.~Font, B.~Fraiman, M.~Gra\~na, C.~A. N\'u\~nez and H.~P. De~Freitas,  {\em {Exploring the landscape of heterotic strings on $T^d$}}, JHEP {\bf 10} (2020) 194 [\href{http://www.arXiv.org/abs/2007.10358}{{\tt 2007.10358}}].

\bibitem{Font:2021uyw}
A.~Font, B.~Fraiman, M.~Gra\~na, C.~A. N\'u\~nez and H.~Parra De~Freitas,  {\em {Exploring the landscape of CHL strings on T$^{d}$}}, JHEP {\bf 08} (2021) 095 [\href{http://www.arXiv.org/abs/2104.07131}{{\tt 2104.07131}}].

\bibitem{Fraiman:2021soq}
B.~Fraiman and H.~P. De~Freitas,  {\em {Symmetry enhancements in 7d heterotic strings}}, JHEP {\bf 10} (2021) 002 [\href{http://www.arXiv.org/abs/2106.08189}{{\tt 2106.08189}}].

\bibitem{Cvetic:2022uuu}
M.~Cveti\v{c}, M.~Dierigl, L.~Lin and H.~Y. Zhang,  {\em {All eight- and nine-dimensional string vacua from junctions}}, Phys. Rev. D {\bf 106} (2022), no.~2, 026007 [\href{http://www.arXiv.org/abs/2203.03644}{{\tt 2203.03644}}].

\bibitem{ParraDeFreitas:2022wnz}
H.~Parra De~Freitas,  {\em {New supersymmetric string moduli spaces from frozen singularities}}, JHEP {\bf 01} (2023) 170 [\href{http://www.arXiv.org/abs/2209.03451}{{\tt 2209.03451}}].

\bibitem{Montero:2022vva}
M.~Montero and H.~Parra~de Freitas,  {\em {New supersymmetric string theories from discrete theta angles}}, JHEP {\bf 01} (2023) 091 [\href{http://www.arXiv.org/abs/2209.03361}{{\tt 2209.03361}}].

\bibitem{Kachru:1995wm}
S.~Kachru and C.~Vafa,  {\em {Exact results for N=2 compactifications of heterotic strings}}, Nucl. Phys. B {\bf 450} (1995) 69--89 [\href{http://www.arXiv.org/abs/hep-th/9505105}{{\tt hep-th/9505105}}].

\bibitem{Israel:2013wwa}
D.~Isra\"el and V.~Thi\'ery,  {\em {Asymmetric Gepner models in type II}}, JHEP {\bf 02} (2014) 011 [\href{http://www.arXiv.org/abs/1310.4116}{{\tt 1310.4116}}].

\bibitem{Hull:2017llx}
C.~Hull, D.~Israel and A.~Sarti,  {\em {Non-geometric Calabi-Yau Backgrounds and K3 automorphisms}}, JHEP {\bf 11} (2017) 084 [\href{http://www.arXiv.org/abs/1710.00853}{{\tt 1710.00853}}].

\bibitem{Gkountoumis:2023fym}
G.~Gkountoumis, C.~Hull, K.~Stemerdink and S.~Vandoren,  {\em {Freely acting orbifolds of type IIB string theory on T$^{5}$}}, JHEP {\bf 08} (2023) 089 [\href{http://www.arXiv.org/abs/2302.09112}{{\tt 2302.09112}}].

\bibitem{Narain:1986qm}
K.~S. Narain, M.~H. Sarmadi and C.~Vafa,  {\em {Asymmetric Orbifolds}}, Nucl. Phys. B {\bf 288} (1987) 551.

\bibitem{Narain:1990mw}
K.~S. Narain, M.~H. Sarmadi and C.~Vafa,  {\em {Asymmetric orbifolds: Path integral and operator formulations}}, Nucl. Phys. B {\bf 356} (1991) 163--207.

\bibitem{Vafa:1996xn}
C.~Vafa,  {\em {Evidence for F theory}}, Nucl. Phys. B {\bf 469} (1996) 403--418 [\href{http://www.arXiv.org/abs/hep-th/9602022}{{\tt hep-th/9602022}}].

\bibitem{Morrison:1996na}
D.~R. Morrison and C.~Vafa,  {\em {Compactifications of F theory on Calabi-Yau threefolds. 1}}, Nucl. Phys. B {\bf 473} (1996) 74--92 [\href{http://www.arXiv.org/abs/hep-th/9602114}{{\tt hep-th/9602114}}].

\bibitem{Morrison:1996pp}
D.~R. Morrison and C.~Vafa,  {\em {Compactifications of F theory on Calabi-Yau threefolds. 2.}}, Nucl. Phys. B {\bf 476} (1996) 437--469 [\href{http://www.arXiv.org/abs/hep-th/9603161}{{\tt hep-th/9603161}}].

\bibitem{Kumar_2010}
V.~Kumar, D.~R. Morrison and W.~Taylor,  {\em {Global aspects of the space of 6D N = 1 supergravities}}, JHEP {\bf 11} (2010) 118 [\href{http://www.arXiv.org/abs/1008.1062}{{\tt 1008.1062}}].

\bibitem{Mizoguchi:2001cp}
S.~Mizoguchi,  {\em {On asymmetric orbifolds and the D = 5 no-modulus supergravity}}, Phys. Lett. B {\bf 523} (2001) 351--356 [\href{http://www.arXiv.org/abs/hep-th/0109193}{{\tt hep-th/0109193}}].

\bibitem{Kumar:2010am}
V.~Kumar, D.~S. Park and W.~Taylor,  {\em {6D supergravity without tensor multiplets}}, JHEP {\bf 04} (2011) 080 [\href{http://www.arXiv.org/abs/1011.0726}{{\tt 1011.0726}}].

\bibitem{Angelantonj:1996mw}
C.~Angelantonj, M.~Bianchi, G.~Pradisi, A.~Sagnotti and Y.~S. Stanev,  {\em {Comments on Gepner models and type I vacua in string theory}}, Phys. Lett. B {\bf 387} (1996) 743--749 [\href{http://www.arXiv.org/abs/hep-th/9607229}{{\tt hep-th/9607229}}].

\bibitem{Morrison:2012np}
D.~R. Morrison and W.~Taylor,  {\em {Classifying bases for 6D F-theory models}}, Central Eur. J. Phys. {\bf 10} (2012) 1072--1088 [\href{http://www.arXiv.org/abs/1201.1943}{{\tt 1201.1943}}].

\bibitem{Seiberg:1996vs}
N.~Seiberg and E.~Witten,  {\em {Comments on string dynamics in six-dimensions}}, Nucl. Phys. B {\bf 471} (1996) 121--134 [\href{http://www.arXiv.org/abs/hep-th/9603003}{{\tt hep-th/9603003}}].

\bibitem{Dolivet:2007sz}
Y.~Dolivet, B.~Julia and C.~Kounnas,  {\em {Magic N=2 supergravities from hyper-free superstrings}}, JHEP {\bf 02} (2008) 097 [\href{http://www.arXiv.org/abs/0712.2867}{{\tt 0712.2867}}].

\bibitem{Kim:2019vuc}
H.-C. Kim, G.~Shiu and C.~Vafa,  {\em {Branes and the Swampland}}, Phys. Rev. D {\bf 100} (2019), no.~6, 066006 [\href{http://www.arXiv.org/abs/1905.08261}{{\tt 1905.08261}}].

\bibitem{Katz:2020ewz}
S.~Katz, H.-C. Kim, H.-C. Tarazi and C.~Vafa,  {\em {Swampland Constraints on 5d $\mathcal{N}=1$ Supergravity}}, JHEP {\bf 07} (2020) 080 [\href{http://www.arXiv.org/abs/2004.14401}{{\tt 2004.14401}}].

\bibitem{Tarazi:2021duw}
H.-C. Tarazi and C.~Vafa,  {\em {On The Finiteness of 6d Supergravity Landscape}}, \href{http://www.arXiv.org/abs/2106.10839}{{\tt 2106.10839}}.

\bibitem{McNamara:2019rup}
J.~McNamara and C.~Vafa,  {\em {Cobordism Classes and the Swampland}}, \href{http://www.arXiv.org/abs/1909.10355}{{\tt 1909.10355}}.

\bibitem{Aspinwall:1996mw}
P.~S. Aspinwall and M.~Gross,  {\em {Heterotic-heterotic string duality and multiple K3 fibrations}}, Phys. Lett. B {\bf 382} (1996) 81--88 [\href{http://www.arXiv.org/abs/hep-th/9602118}{{\tt hep-th/9602118}}].

\bibitem{Ganor:1996mu}
O.~J. Ganor and A.~Hanany,  {\em {Small E(8) instantons and tensionless noncritical strings}}, Nucl. Phys. B {\bf 474} (1996) 122--140 [\href{http://www.arXiv.org/abs/hep-th/9602120}{{\tt hep-th/9602120}}].

\bibitem{Blumenhagen:2016rof}
R.~Blumenhagen, M.~Fuchs and E.~Plauschinn,  {\em {The Asymmetric CFT Landscape in D=4,6,8 with Extended Supersymmetry}}, Fortsch. Phys. {\bf 65} (2017), no.~3-4, 1700006 [\href{http://www.arXiv.org/abs/1611.04617}{{\tt 1611.04617}}].

\bibitem{Candelas:2016fdy}
P.~Candelas, A.~Constantin and C.~Mishra,  {\em {Calabi-Yau Threefolds with Small Hodge Numbers}}, Fortsch. Phys. {\bf 66} (2018), no.~6, 1800029 [\href{http://www.arXiv.org/abs/1602.06303}{{\tt 1602.06303}}].

\bibitem{Dabholkar:1998kv}
A.~Dabholkar and J.~A. Harvey,  {\em {String islands}}, JHEP {\bf 02} (1999) 006 [\href{http://www.arXiv.org/abs/hep-th/9809122}{{\tt hep-th/9809122}}].

\bibitem{Carter1970}
R.~Carter, {\em Conjugacy classes in the Weyl group}, pp.~297--318.
\newblock Springer Berlin Heidelberg, Berlin, Heidelberg, 1970.

\bibitem{Bouwknegt:1988hn}
P.~Bouwknegt,  {\em {Lie Algebra Automorphisms, the Weyl Group and Tables of Shift Vectors}}, J. Math. Phys. {\bf 30} (1989) 571.

\bibitem{Lee:2019wij}
S.-J. Lee, W.~Lerche and T.~Weigand,  {\em {Emergent strings from infinite distance limits}}, JHEP {\bf 02} (2022) 190 [\href{http://www.arXiv.org/abs/1910.01135}{{\tt 1910.01135}}].

\bibitem{Lee:2019xtm}
S.-J. Lee, W.~Lerche and T.~Weigand,  {\em {Emergent strings, duality and weak coupling limits for two-form fields}}, JHEP {\bf 02} (2022) 096 [\href{http://www.arXiv.org/abs/1904.06344}{{\tt 1904.06344}}].

\bibitem{Dixon:1985jw}
L.~J. Dixon, J.~A. Harvey, C.~Vafa and E.~Witten,  {\em {Strings on Orbifolds}}, Nucl. Phys. B {\bf 261} (1985) 678--686.

\bibitem{Dixon:1986jc}
L.~J. Dixon, J.~A. Harvey, C.~Vafa and E.~Witten,  {\em {Strings on Orbifolds. 2.}}, Nucl. Phys. B {\bf 274} (1986) 285--314.

\bibitem{Narain:1985jj}
K.~S. Narain,  {\em {New Heterotic String Theories in Uncompactified Dimensions \ensuremath{<} 10}}, Phys. Lett. B {\bf 169} (1986) 41--46.

\bibitem{Narain:1986am}
K.~S. Narain, M.~H. Sarmadi and E.~Witten,  {\em {A Note on Toroidal Compactification of Heterotic String Theory}}, Nucl. Phys. B {\bf 279} (1987) 369--379.

\bibitem{Harvey:2017rko}
J.~A. Harvey and G.~W. Moore,  {\em {An Uplifting Discussion of T-Duality}}, JHEP {\bf 05} (2018) 145 [\href{http://www.arXiv.org/abs/1707.08888}{{\tt 1707.08888}}].

\bibitem{Vafa:1986wx}
C.~Vafa,  {\em {Modular Invariance and Discrete Torsion on Orbifolds}}, Nucl. Phys. B {\bf 273} (1986) 592--606.

\bibitem{Erler:1993zy}
J.~Erler,  {\em {Anomaly cancellation in six-dimensions}}, J. Math. Phys. {\bf 35} (1994) 1819--1833 [\href{http://www.arXiv.org/abs/hep-th/9304104}{{\tt hep-th/9304104}}].

\bibitem{Aldazabal:1997wi}
G.~Aldazabal, A.~Font, L.~E. Ibanez, A.~M. Uranga and G.~Violero,  {\em {Nonperturbative heterotic D = 6, D = 4, N=1 orbifold vacua}}, Nucl. Phys. B {\bf 519} (1998) 239--281 [\href{http://www.arXiv.org/abs/hep-th/9706158}{{\tt hep-th/9706158}}].

\bibitem{Sagnotti:1992qw}
A.~Sagnotti,  {\em {A Note on the {G}reen-{S}chwarz mechanism in open string theories}}, Phys. Lett. {\bf B294} (1992) 196--203
[\href{http://www.arXiv.org/abs/hep-th/9210127}{{\tt hep-th/9210127}}].

\bibitem{Kumar:2010ru}
V.~Kumar, D.~R. Morrison and W.~Taylor,  {\em Global aspects of the space of 6{D} $ \mathcal{N} = 1 $ supergravities}, Journal of High Energy Physics {\bf 2010} (Nov, 2010).

\bibitem{Seiberg:2011dr}
N.~Seiberg and W.~Taylor,  {\em {Charge Lattices and Consistency of 6D Supergravity}}, JHEP {\bf 06} (2011) 001 [\href{http://www.arXiv.org/abs/1103.0019}{{\tt 1103.0019}}].

\bibitem{Banks:2010zn}
T.~Banks and N.~Seiberg,  {\em {Symmetries and Strings in Field Theory and Gravity}}, Phys. Rev. D {\bf 83} (2011) 084019 [\href{http://www.arXiv.org/abs/1011.5120}{{\tt 1011.5120}}].

\bibitem{Polchinski:2003bq}
J.~Polchinski,  {\em {Monopoles, duality, and string theory}}, Int. J. Mod. Phys. A {\bf 19S1} (2004) 145--156 [\href{http://www.arXiv.org/abs/hep-th/0304042}{{\tt hep-th/0304042}}].

\bibitem{Monnier:2018nfs}
S.~Monnier and G.~W. Moore,  {\em {Remarks on the Green\textendash{}Schwarz Terms of Six-Dimensional Supergravity Theories}}, Commun. Math. Phys. {\bf 372} (2019), no.~3, 963--1025 [\href{http://www.arXiv.org/abs/1808.01334}{{\tt 1808.01334}}].

\bibitem{Chaudhuri:1995fk}
S.~Chaudhuri, G.~Hockney and J.~D. Lykken,  {\em {Maximally supersymmetric string theories in D \ensuremath{<} 10}}, Phys. Rev. Lett. {\bf 75} (1995) 2264--2267 [\href{http://www.arXiv.org/abs/hep-th/9505054}{{\tt hep-th/9505054}}].

\bibitem{Chaudhuri:1995bf}
S.~Chaudhuri and J.~Polchinski,  {\em {Moduli space of CHL strings}}, Phys. Rev. D {\bf 52} (1995) 7168--7173 [\href{http://www.arXiv.org/abs/hep-th/9506048}{{\tt hep-th/9506048}}].

\bibitem{Bershadsky:1996nh}
M.~Bershadsky, K.~A. Intriligator, S.~Kachru, D.~R. Morrison, V.~Sadov and C.~Vafa,  {\em {Geometric singularities and enhanced gauge symmetries}}, Nucl. Phys. B {\bf 481} (1996) 215--252 [\href{http://www.arXiv.org/abs/hep-th/9605200}{{\tt hep-th/9605200}}].

\bibitem{Kawai:1986va}
H.~Kawai, D.~C. Lewellen and S.~H.~H. Tye,  {\em {Construction of Four-Dimensional Fermionic String Models}}, Phys. Rev. Lett. {\bf 57} (1986) 1832 [Erratum: Phys.Rev.Lett. 58, 429 (1987)].

\bibitem{Lerche:1986cx}
W.~Lerche, D.~Lust and A.~N. Schellekens,  {\em {Chiral Four-Dimensional Heterotic Strings from Selfdual Lattices}}, Nucl. Phys. B {\bf 287} (1987) 477.

\bibitem{Kawai:1986ah}
H.~Kawai, D.~C. Lewellen and S.~H.~H. Tye,  {\em {Construction of Fermionic String Models in Four-Dimensions}}, Nucl. Phys. B {\bf 288} (1987) 1.

\bibitem{Antoniadis:1986rn}
I.~Antoniadis, C.~P. Bachas and C.~Kounnas,  {\em {Four-Dimensional Superstrings}}, Nucl. Phys. B {\bf 289} (1987) 87.

\bibitem{Kiritsis:2008mu}
E.~Kiritsis, M.~Lennek and B.~Schellekens,  {\em {Free Fermion Orientifolds}}, JHEP {\bf 02} (2009) 030 [\href{http://www.arXiv.org/abs/0811.0515}{{\tt 0811.0515}}].

\bibitem{Anastasopoulos:2009kj}
P.~Anastasopoulos, M.~Bianchi, J.~F. Morales and G.~Pradisi,  {\em {(Unoriented) T-folds with few T's}}, JHEP {\bf 06} (2009) 032 [\href{http://www.arXiv.org/abs/0901.0113}{{\tt 0901.0113}}].

\bibitem{Farragi:2023}
A.~Farragi, G.~Nibbelink and B.~Percival,  {Checking the DJK Model: D = 4 Type II Asymmetric Orbifold, Unpublished}.

\bibitem{Solr-2759686}
.~J.~. Fuchs, Jürgen, {\em Symmetries, lie algebras and representations : a graduate course for physicists}.
\newblock Cambridge monographs on mathematical physics. Cambridge University Press, Cambridge, U.K. ; New York, NY, USA, 1997.

\end{thebibliography}\endgroup

\end{document}